\documentclass[10pt,a4paper]{article}
\usepackage[utf8]{inputenc}
\usepackage{indentfirst}
\usepackage{float}
\usepackage{xcolor}
\usepackage[acronym]{glossaries}
\usepackage{multirow}
\usepackage{array}
\usepackage{tabularx}
\usepackage{chngpage}
\usepackage{amssymb}

\makeglossaries
\newacronym{GNSS}{GNSS}{Global Navigation Satellite Systems}
\newacronym{IDS}{IDS}{Intrusion Detection System}
\newacronym{TESLA}{TESLA}{Timed Efficient Stream Loss-tolerant Authentication}
\newacronym{UTM}{UTM}{UAV Traffic Management}
\newacronym{GCS}{GCS}{Ground Control Station}
\newacronym{GS}{GS}{Ground Station}
\newacronym{UAV}{UAV}{Unmanned Aerial Vehicle}
\newacronym[plural=UASs,firstplural=Unmanned Aerial Systems (UAS)]{UAS}{UAS}{Unmanned Aerial System}
\newacronym{ADS-B}{ADS-B}{Automatic Dependent Surveillance-Broadcast}
\newacronym{FFX}{FFX}{Format-preserving, Feistel-based Encryption}
\newacronym{VLOS}{VLOS}{Visual Line Of sight}
\newacronym{BVLOS}{BVLOS}{Beyond the Visual Line Of Sight}
\newacronym{IoT}{IoT}{Internet of Things}
\newacronym{IoD}{IoD}{Internet of Drones}
\newacronym{NTSC}{NTSC}{National Television System Commitee}
\newacronym{PAL}{PAL}{Phase Alternating Line}
\newacronym{SECAM}{SECAM}{Sequential Color and Memory}
\newacronym{LFSR}{LFSR}{Linear Feedback Shift Register}
\newacronym{AES}{AES}{Advanced Encryption Standard}
\newacronym{DOS}{DOS}{Denial of Service}
\newacronym{MITM}{MITM}{Man-In-The-Middle}
\newacronym{MANET}{MANET}{Mobile Ad-Hoc Network}
\newacronym[plural=VANETs,firstplural=Vehicular Ad-Hoc Networks (VANETs)]{VANET}{VANET}{Vehicular Ad-Hoc Network}
\newacronym{FANET}{FANET}{Flying Ad-Hoc Network}
\newacronym{AP}{AP}{Access Point}
\newacronym{MIMO}{MIMO}{Multiple-Input Multiple-Output}
\newacronym{IRS}{IRS}{Intelligent Reflecting Surface}
\newacronym{FHSS}{FHSS}{Frequency Hopping Spread Spectrum}
\newacronym{FTP}{FTP}{File Transfer Protocol}
\newacronym{TLS}{TLS}{Transport Layer Protocol}
\newacronym{DH}{DH}{Diffie-Hellman}
\newacronym{RSA}{RSA}{Rivest Shamir Adelman}
\newacronym{AKA}{AKA}{Authenticated Key Agreement}
\newacronym{AKE}{AKE}{Authenticated Key Exchange}
\newacronym{RF}{RF}{Radio Frequency}
\newacronym{ML}{ML}{Machine Learning}
\newacronym{FAA}{FAA}{Federal Aviation Administration}
\newacronym{NBAA}{NBAA}{National Business Aviation Association}
\newacronym{ICAO}{ICAO}{International Civil Aviation Organization}
\newacronym{PUF}{PUF}{Physical Unclonable Function}
\newacronym{ECC}{ECC}{Elliptic Curve Cryptography}
\newacronym{AEAD}{AEAD}{Authenticated Encryption with Associated Data}
\newacronym{CPU}{CPU}{Central Processing Unit}
\newacronym{NIST}{NIST}{National Institute of Standards and Technology}
\newacronym{ATC}{ATC}{Air Traffic Control}
\newacronym{ATM}{ATM}{Air Traffic Management}
\newacronym{CNS}{CNS}{Communications, Navigation and Surveillance}
\newacronym{EASA}{EASA}{European Aviation Safety Agency}
\newacronym{WSN}{WSN}{Wireless Sensor Network}
\newacronym{LWC}{LWC}{Lightweight Cryptography}
\newacronym{MAC}{MAC}{Message Authentication Code}
\newacronym{DES}{DES}{Data Encryption Standard}
\newacronym{DHE}{DHE}{Diffie-Hellman Ephemeral}
\newacronym{ECDHE}{ECDHE}{Elliptic Curve Diffie-Hellman Ephemeral}
\newacronym{ECDSA}{ECDSA}{Elliptic Curve Digital Signature Algorithm}
\newacronym{EdDSA}{EdDSA}{Edwards Curve Digital Signature Algorithm}
\newacronym{RSA-PSS}{RSA-PSS}{RSA with Probabilistic Signature Scheme}
\newacronym{KEM}{KEM}{Key Encapsulation Mechanism}
\newacronym{DEM}{DEM}{Data Encapsulation Mechanism}
\newacronym{SPOF}{SPOF}{Single Point Of Failure}
\newacronym{PQC}{PQC}{Post-Quantum Cryptography}
\newacronym{HMAC}{HMAC}{Hash-based Message Authentication Code}
\newacronym{ARC}{ARC}{Aviation Rukemaking Committee}
\newacronym{SDR}{SDR}{Software Defined Radio}
\newacronym{OTP}{OTP}{One-Time-Pad}
\newacronym{XOR}{XOR}{Exclusive Or}

\usepackage{graphicx}
\graphicspath{ {./images/} }

\usepackage[
backend=biber,
style=numeric,
maxbibnames=99
]{biblatex}

\toks0\expandafter{\biburlsetup}
\edef\biburlsetup{\the\toks0 \Urlmuskip =0mu\relax}

\title{UAV Traffic Management : A Survey On Communication Security }
\author{Ridwane Aissaoui \and Jean-Christophe Deneuville \and Christophe Guerber \and Alain Pirovano}

\date{
\text{ENAC, French Civil Aviation University, Toulouse, France}\\[5mm]
\texttt{<first.last@enac.fr>} \\[12mm]
\today}

\begin{document}

\sloppy

\maketitle

\begin{abstract}

\glspl{UAS} have a wide variety of applications, and their development in terms of capabilities is continuously evolving. Many missions performed by an \gls{UAV} require flying in public airspace. This requires very high safety standards, similar to those mandatory in commercial civil aviation. A safe \gls{UTM} requires several communication links between aircraft, their pilots and UTM systems. The integrity of these communication links is critical for the safety of operations. Several security requirements also have to be met on each of these links. Unfortunately, current cryptographic standards used over the internet are most often not suitable to UAS due to their limited resources and dynamic nature. This survey discusses the security required for every communication link in order to enable a safe traffic management. Research works focusing on the security of communication links using cryptographic primitives are then presented and discussed. Authentication protocols developed for UAVs or other constrained systems are compared and evaluated as solutions for UAS security. Symmetrical alternatives to the AES algorithm are also presented. Works to secure current UTM protocols such as ADS-B and RemoteID are discussed. The analysis reveals a need for the development of a complete secure architecture able to provide authentication and integrity to external systems (other aircraft, UTM systems...).

\end{abstract}

\emph{Keywords : UAV, Drone, UAS, UTM, CNS, Information Security, Cryptography, Authentication, Safety, ADS-B, RemoteID, IoD}

\section{Introduction }

\subsection{Unmanned aerial systems applications and\newline traffic management}

The most common application for a civil \gls{UAV} in 2022 is the recreational use. However they have proven to be critical in operations that humans cannot carry out in a safe and timely efficient manner \cite{shakhatreh_unmanned_2019}. The number of \gls{UAV}s around the world grows by 13\% each year, and a lot of research focuses on improving their operating capabilities. Their performance is continuously improving and they are the best solution for a growing number of applications. For infrastructure monitoring, areas scanning as well as urgent delivery services and other applications, they are already the most relevant and cost-efficient solution. They can also be used for agriculture by monitoring and spraying the fields, for transport to help limiting the congestion in city centers, for the surveillance of areas where security camera are unusable or more expensive, for telecommunication purposes and for media and entertainment as cheap aerial cameras or to create new shows. They also can play a major role in smart cities and be used in \gls{IoT} systems or \gls{WSN}~\cite{de_freitas_uav_2010}.

A \gls{UAS} includes all the parts used to operate a \gls{UAV} and their communication means. The simplest ones are comprised of a \gls{UAV} and a \gls{GCS}. Many different actors are taking part in advanced \gls{UAS} such as \gls{UTM} systems and intermediate ground stations managing communication between different \gls{UAV}s and their end users. \gls{UAS} characteristics cause most communication links to be wireless. As shown in Figure~\ref{fig:utm}, three main communication axis can be found in a \gls{UAS}. There are links between any \gls{UAV} and a \gls{GCS}, where command, telemetry, video and other mission specific data is transmitted. These links can be physically or logically separated, as these different types of data are not always sent on the same channel. A second axis exists between any \gls{UAS} and the \gls{UTM} systems when flying in controlled airspace. From the drone or the \gls{GCS}, there is telemetry information sent to the \gls{UTM} systems, which uses it to monitor the traffic and organize the airspace. And from the \gls{UTM} systems, there are emergency geofencing zones broadcast, and, depending on the level of authority of the \gls{UTM}, trajectory suggestions or direct trajectory modifications sent to a specific \gls{UAV} or \gls{GCS}. The third type of communications are between two \gls{UAV}s. They can share environmental information with one another, or be used as routers to transmit data to a remote \gls{GCS} or the \gls{UTM}. Depending on the sensitivity of the information in each transmission, the security goals will be different.

\subsection{\gls{UAS} security issues}

In Europe the regulatory framework strongly limits \gls{UAS} operations. Most \gls{UAV} operations require or gain from \gls{BVLOS} flights. This type of flight requires a sustained communication link between the \gls{GCS} and the \gls{UAV}, as the pilot has to be able to take control of the \gls{UAV} at any time during its flight. Depending on the category of \gls{UAV} as defined by the \gls{EASA}, flights are more or less restricted. The open, specific and certified categories separate \gls{UAV}s by their characteristics, and are allowed to perform different civil risk level operations~\cite{noauthor_civil_nodate}. Drones in the open category are not allowed to perform \gls{BVLOS} flights. The specific category allows \gls{BVLOS} flights up to 1 km (2 with airspace observers) and only for \gls{UAV}s under 25kg~\cite{noauthor_easy_nodate}. These numbers are more restricted when flying over densely populated areas. This severely limits the operations. The certified category should broaden the limits but as the certification process is slow and rigorous, it will take time to complete for manufacturers. As most applications need to break these limits, each \gls{UAV} used will need to be certified or the specific category will need to be modified. For information security, no rules have been announced by the \gls{EASA}, and the \gls{FAA} has been advised by the \gls{ARC}~\cite{aviaion_rulemaking_committee_unmanned_2022} to build a working group to address the issue. 

\gls{UAS} are targeted by numerous cyberthreats. They can contain sensitive and private information, they can transport payloads of large financial or human value (high end goods or medical supplies) which can be recovered by attackers if the \gls{UAS} is compromised. They also pose a physical threat to the public with their kinetic energy. It is therefore important to protect \gls{UAS}s against cyber attacks. Most of these threats rely on compromising the communication links in the \gls{UAS}.

\subsection{\gls{UTM} safety and security}

\gls{UAS} will need to meet a great number of security requirements in order to operate over populated areas. Among those a great deal will be focused on communication security. They will need to be included in a controlled airspace, managed by \gls{UTM} systems synchronised with existing \gls{ATM} system~\cite{mccarthy_fundamental_2020}. Most of the information exchanged by \gls{UAS} and \gls{UTM} systems is sensitive, if not critical for safe operations. A secure \gls{UTM} requires solutions to protect communications. The research field has anticipated this and several works have focused on securing \gls{UAS} communications~\cite{hayat_survey_2016, shafique_survey_2021, sharma_communication_2020}.

The \gls{ICAO} issued standard definitions for security and safety~\cite{chen_documents_2017}, as being:
\begin{center}
    \begin{minipage}{.9\textwidth}
    ``\emph{the state in which risks associated with aviation activities, related to, or in direct support of the operation of aircraft, are reduced and controlled to an acceptable level. Security aims to safeguard civil aviation against acts of unlawful interference. The objective is achieved by a combination of measures and human and material resources.}''
    \end{minipage}
\end{center}
Despite being two separate notions in the regulatory framework, the safety of aerial operations cannot be guaranteed without security measures. The concept of security for safety has emerged to cover the security solutions involved in reaching the safety objectives. In this paper, this concept is simply addressed as safety.

Unfortunately, defending against cyber attacks can require computation power, memory space, bandwidth, sizeable software or dedicated hardware. \gls{UAS} are highly constrained systems, and their operability is partially based on the cost-efficiency of their usage. The defense mechanisms have to be adapted and optimized to be efficiently implemented. More specifically, information security in communications is mostly based on cryptography. Specific lightweight cryptographic primitives are developed for constrained systems. These lightweight primitives aim to provide sufficient security while requiring lower performance. 

The reminder of the paper is organized as follows: Communication links' vulnerabilities that impact the safety of \gls{UTM} are discussed in Section~\ref{sec:vuln}. The importance of the level of security on each link is evaluated from a \gls{UTM} point of view. \gls{GNSS} vulnerabilities are omitted as they are considered out of the communication means included in \gls{UTM} systems. \gls{GNSS} vulnerabilities and solutions are studied by the GPS, Galileo and other \gls{GNSS} developer teams. In Section~\ref{sec:crypto} this paper presents cryptographic methods that are relevant for use in a \gls{UAS} environment for two-way communications. We discuss lightweight encryption solutions with different characteristics from the \gls{UAS} field of work, the \gls{IoT} and other constrained devices cryptographic propositions. Asymmetric encryption (Section \ref{subsec:auth}) and symmetric encryption (Section \ref{subsec:sym}) are discussed and compared for performance and security level. Section~\ref{sec:broadcast} focuses on security solutions for broadcast communications, including signature schemes. Section~\ref{sec:challenges} discusses the open issues and proposes some ideas for future research in the field of \gls{UTM} security.

\section{Context}

\subsection{Types of \gls{UAV}s}

\gls{UAV}s are generally categorized by weight. The physical threat they pose for the public is largely dependent on the kinetic energy they hold when flying over populated areas. Larger \gls{UAV}s embark powerful hardware, as the energy consumption, weight and price of \gls{CPU}s is negligible compared to the advantages they bring. These larger \gls{UAV}s are able to use internet protocols to provide information security to their users. For small \gls{UAV}s, these solutions are not cost or energy efficient, as the performance requirements severely impact their cost and range. The use of smaller computational units is preferred but their lower performance is insufficient for all the computation required by the security features of internet protocols. Similarly, the bandwidth of the communication links is also a limiting factor.

\begin{table}[H]
    \centering
    \begin{tabular}{ | c | c | c | c | c | } 
      \hline
      Category & Weight & Altitude & Range & Payload \\ 
      \hline
      Nano & \(<\) 0.2 kg & \(<\) 90 m & \(<\) 90 m & \(<\) 0.2 kg \\ 
      \hline
      Micro & \(<\) 2 kg & \(<\) 90 m & \(<\) 5 km & \(<\) 0.5 kg \\ 
      \hline
      Mini & \(<\) 20 kg & \(<\) 900 m & \(<\) 25 km & \(<\) 10 kg \\
      \hline
      Small & \(<\) 150 kg & \(<\) 1500 m & \(<\) 100 km & \(<\) 50 kg \\
      \hline
      Tactical & \(>\) 150 kg & \(>\) 1500 m & \(>\) 100 km & \(>\) 50 kg \\
      \hline
    \end{tabular}
    \caption{Drone categories as proposed in \cite{nassi_sok_2019}.}
    \label{table:dronecat}
\end{table}

Only \gls{UAV}s in the categories Mini and smaller in Table~\ref{table:dronecat} are considered to be limited in performance. This survey focuses on civil applications using these smaller \gls{UAV}s. This corresponds to the open and part of the specific \gls{UAV} categories of \gls{EASA} regulations. The improvement of information security (plus operational safety and other security issues) is needed for the drones in the open category to operate in different areas. These improvements will allow A1, A2 and A3 \gls{UAV}s (as defined in Table~\ref{table:open}) to operate in \gls{BVLOS} and over populated areas.

\begin{table}[H]
    \begin{adjustwidth}[]{-3cm}{-3cm}
    \centering
    \resizebox{1.4\textwidth}{!}{
        \begin{tabularx}{1.5\textwidth}{ | p{1.5cm} | p{1.7cm} | p{2.5cm} | p{3cm} | X | p{1.7cm} | }  
          \hline
          \centering UAS & \multicolumn{2}{c|}{Operation} & \multicolumn{3}{c|}{Drone Operator/Pilot}\\
          \hline
          \centering Max Weight  &\centering Subcategory  &\centering Operational Restrictions  &\centering Drone Operator Registration  &\centering Remote Pilot Competence  & \multicolumn{1}{p{1.7cm}|}{\centering Remote Pilot Minimum Age} \\
          \hline
          \begin{center}\(<\)250g \end{center} & \multirow{2}{\hsize}{\begin{center} A1 (can also fly in subcategory A3) \end{center}} & \multirow{2}{\hsize}{\begin{center}- No flight expected over uninvolved people (if it happens, overflight should be minimised)\newline \newline - No flight over assemblies of people \end{center}} &\begin{center} No, unless camera / sensor on board and the drone is not a toy \end{center} &\begin{center} - No training Required \end{center} & \begin{center}No minimum age\end{center}\\
          \cline{1-1} \cline{4-6}
          \begin{center}\(<\)500g \end{center} &&& \begin{center} Yes \end{center} & \begin{center}- Read carefully the user manual \newline \newline - Complete the training and pass the exam defined by your national competent authority or have a 'Proof of completion for online training' for A1/A3 'open' subcategory \end{center} & \begin{center}16+\end{center} \\
          \hline
          \begin{center}\(<\)2kg \end{center} & \begin{center} A2 (can also fly in subcategory A3) \end{center}&\begin{center} - No flying over uninvolved people \newline \newline - Keep a horizontal distance of 50 m from uninvolved people \end{center} & \begin{center}Yes \end{center} &\begin{center} - Read carefully the user manual \newline \newline - Complete the training and pass the exam defined by your national competent authority or have a 'remote pilot certificate of competency' for A2 'open' subcategory \end{center} & \begin{center}16+\end{center} \\
          \hline
          \begin{center}\(<\)25kg \end{center} & \begin{center}A3 \end{center} & \begin{center}- Do not fly near or over people \newline \newline - Fly at least 150m away from residential, commercial or industrial areas \end{center} &\begin{center} Yes \end{center} &\begin{center} - Read carefully the user manual \newline\newline - Complete the training and pass the exam defined by your national competent authority or have a 'Proof of completion for online training' for A1/A3 'open' subcategory \end{center} &\begin{center} 16+ \end{center} \\
          \hline
        \end{tabularx}
}
    \end{adjustwidth}
        \caption{\gls{EASA} open category~\cite{noauthor_open_nodate}.}
        \label{table:open}
\end{table}

\subsection{\gls{UAS} networks }

A \gls{UAS} with multiple \gls{UAV}s is organised as either a cellular network or a \gls{FANET}. Ad-Hoc networks have specific security solutions that are adapted to their unique properties. The specific security requirements and early security solutions of Ad-Hoc networks are detailed in \cite{zhou_securing_1999}. A \gls{FANET} is a specific \gls{MANET} with very high node mobility and speed, a lower node density and a specific mobility model \cite{noauthor_manet_nodate}. Specific communication architectures have been proposed for \gls{FANET}s, with slight variations compared to \gls{MANET} solutions. These architectures have been surveyed in \cite{chriki_fanet_2019}.

\begin{figure}[H]
    \centering
    \includegraphics[width=0.6\textwidth]{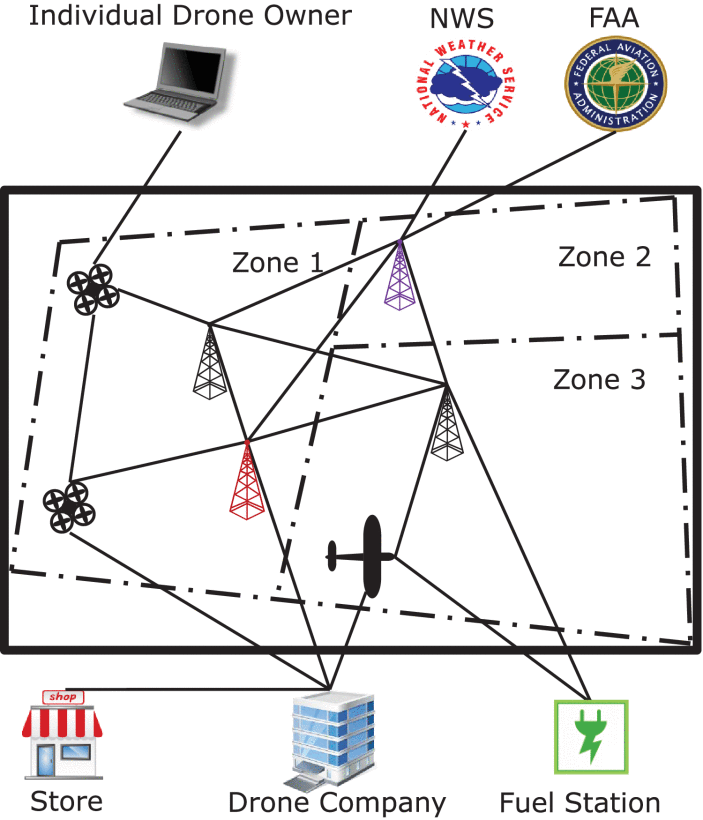}
    \caption{\gls{IoD} architecture organization \cite{gharibi_internet_2016}.}
    \label{fig:iod}
\end{figure}

A popular proposition of a \gls{UAS} control architecture is the \gls{IoD} proposed by Gharibi, Boutaba and Waslander~\cite{gharibi_internet_2016}. It is designed for the coordination of \gls{UAV}s into a controlled airspace. The \gls{IoD} includes an \gls{UTM} system using \gls{UAV}s as aerial relays, thus building a \gls{FANET} within the system (see Figure~\ref{fig:iod}). The \gls{IoD} structure is based around network concepts from the \gls{ATC}, the cellular network and the Internet. 

The \gls{IoD} structure can be deployed for a large variety of applications such as smart city surveillance, \gls{WSN}, business-to-business...~\cite{abualigah_applications_2021}. However the system lacks interoperability with \gls{ATM} systems. This is explained by the non invasion of controlled airspace from \gls{UAS} not in direct communication with the \gls{ATC}. \gls{ATC} remains in charge of aircraft separation for both manned and unmanned systems~\cite{gharibi_internet_2016}.

\subsection{UAS traffic management organisation }
\label{subsec:utm}

In this survey, \gls{UTM} is considered as defined by the \gls{ICAO} framework~\cite{icao_unmanned_2019}:
\begin{center}
    \begin{minipage}{.9\textwidth}
    ``\emph{\gls{UTM} is a special aspect of air traffic management which manages \gls{UAS} operations safely, economically and efficiently through the provision of facilities and a seamless set of services in collaboration with all parties and involving airborne and ground-based functions. The \gls{UTM} system provides \gls{UTM} through the collaborative integration of humans, information, technology, facilities and services, supported by air, ground or space-based \gls{CNS}.}''
    \end{minipage}
\end{center}

\begin{figure}[ht]
    \begin{adjustwidth}[]{-8cm}{-8cm}
    \centering
    \includegraphics[width=1.5\textwidth]{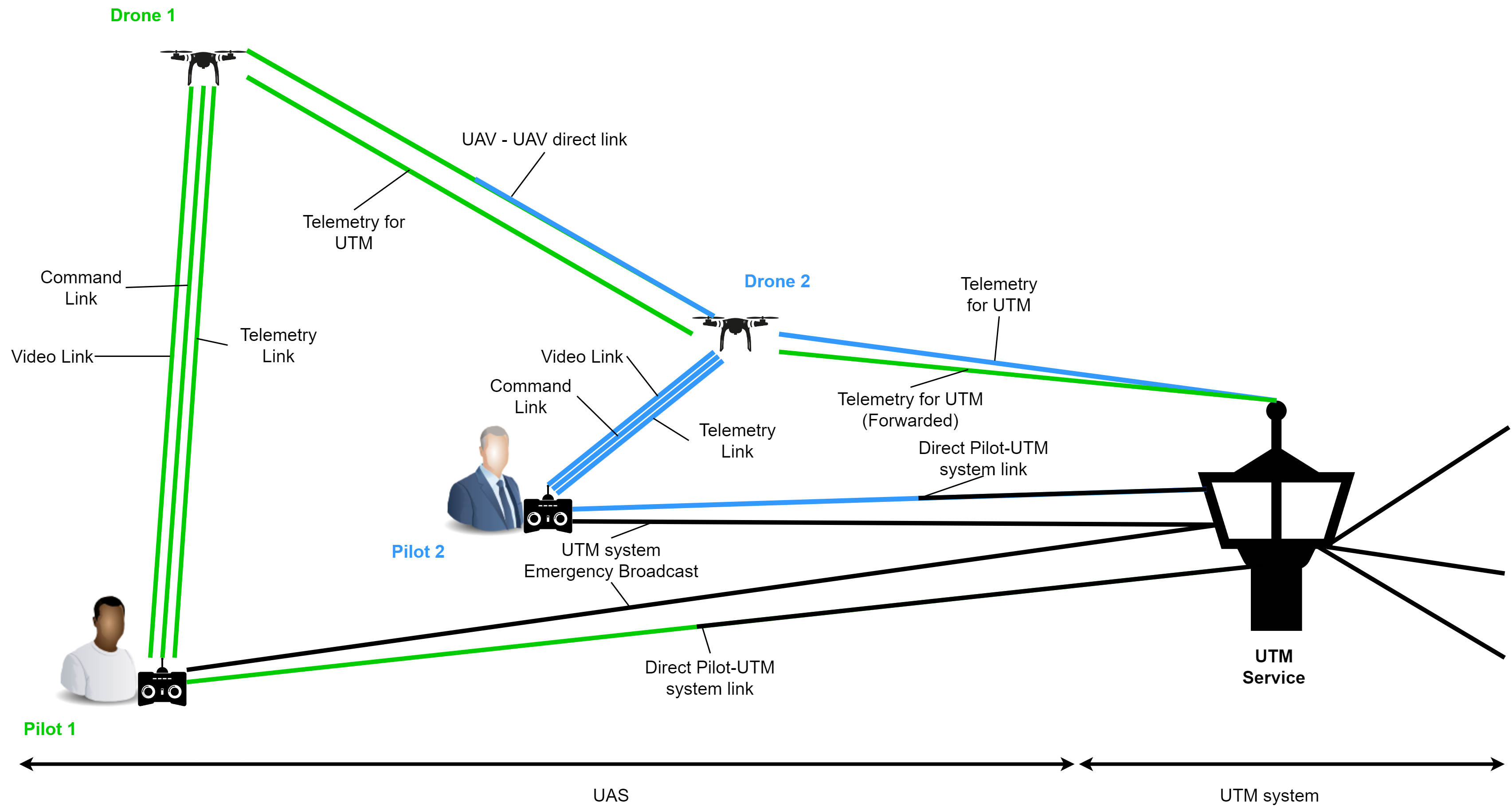}
    \caption{\gls{UAS} with \gls{UTM} system organisation.}
    \label{fig:utm}
    \end{adjustwidth}
\end{figure}

In Figure~\ref{fig:utm}, only the system providing the \gls{UTM} flight control service is represented as it is the part of an \gls{UTM} system that communicates with the \gls{UAS}.

The \gls{ICAO} states that \gls{UTM} system information protocols and interfaces play a key role in the safe integration of \gls{UAS}s into public airspace. \cite{icao_unmanned_2019}~precises that it is necessary to develop minimum performance and interoperability standards for communications protocols for :
\begin{itemize}
    \item C2 Link between \gls{GCS} and \gls{UAV}s
    \item \gls{UAV} to \gls{UAV} communications
    \item Communications link between \gls{UAV}s and other airspace users (e.g. manned aircraft), as necessary
    \item Communications between remote pilots/\gls{GCS}s and the \gls{UTM} and \gls{ATM} systems.
\end{itemize}

One of the pillars of traffic management systems is the \gls{CNS} concept. It includes the development of communication systems. These systems enable a safe navigation for aircraft. And the traffic is observed by surveillance systems, to ensure safe operations.

This survey focuses on the security requirements for \gls{UAS}s communications to enable a safe traffic management. \gls{BVLOS} operations are used for security evaluation as they represent the focus point of \gls{UTM}~\cite{mccarthy_fundamental_2020} research. This means that the video link is critical for \gls{UAV} control, alongside the command link. Safe traffic management requires different levels of information security in case of~:
\begin{itemize}
    \item \gls{UTM} system broadcast for emergency geofencing or modification of airspace organization requires authentication of the \gls{UTM} system and integrity while keeping a high availability level.
    \item \gls{UTM} system direct link to \gls{UAS} (\gls{GCS} or \gls{UAV}) is used for real-time traffic control, and is critical for safety as it modifies \gls{UAV} paths. Authentication and integrity is needed.
    \item Commands from the \gls{GCS} to the \gls{UAV} require authentication and integrity.
    \item Video from the \gls{UAV} to the \gls{GCS} is critical for safe \gls{UAV} control from the pilot. It has the same requirements as the command link.
    \item Telemetry from the \gls{UAV} is transmitted to the \gls{GCS} in its entirety and is used for safe operation of the \gls{UAS}. For \gls{BVLOS} flights, this information is critical for the safety of the operation as it is important for the surveillance of the flight. It requires authentication and integrity.
    \item A subpart of telemetry data is broadcast from the \gls{UAV} to neighboring \gls{UAV}s (for detect and avoid systems) and to the \gls{UTM} system for real-time traffic surveillance. Authentication and integrity is needed. 
\end{itemize}

For every one of these links, confidentiality is a privacy concern, and does not directly impact safety. However it is important to consider it when designing a system, as privacy issues have to be addressed. Moreover, the safety of the system can be indirectly impacted by the lack of confidentiality (i.e. intercepting a drone by anticipating its trajectory). The same reasoning exists for the non-repudiation property, which is a traceability issue. This property is necessary to allow authorities to investigate previous incidents, but does not directly impact air traffic safety. Confidentiality and non-repudiation can ensue from schemes that provide authentication and integrity. Going further into this survey, confidentiality and non-repudiation will be discussed when addressed by research works, but not considered as a major issue regarding \gls{UTM} safety.

\subsection{Cryptography for information security}

\subsubsection{Main principles }

Information security is based around three main concepts~: 
\begin{itemize}
    \item Confidentiality ensures that no illegitimate user can access the information
    \item Integrity verifies that the information has not been modified during the lifetime of the data
    \item Availability is the property allowing legitimate use of the information
\end{itemize}

There are several ways to add security to a communication link. As the links are wireless and the network topology is dynamic in a \gls{UAS}, it excludes the possibility of physically protecting the links. Using cryptography is the best way to secure a communication link with those specifications.

To ensure information security, cryptographic techniques have been researched since the antiquity. The long history of cryptography has produced many proven solutions for information security. In cryptography, there are four properties that ensure data security : 
\begin{itemize}
    \item Confidentiality ensures that only a legitimate user can access the information
    \item Integrity verifies that the information has not been modified during the lifetime of the data
    \item Authentication provides trust in the identity of the parties sharing information
    \item Non-Repudiation prevents the source of the information from denying being the source
\end{itemize}

If good practices and standards are correctly applied in the implementation of cryptographic methods, information can be theoretically confidential, authenticated, non repudiable and its integrity ensured. The Kerckhoff principle states that the security of a scheme should only rely on the secrecy of a randomly chosen parameter (called a key), rather than on the secrecy of the algorithm. Actually, all the algorithms of the scheme should be public, and only the knowledge of the secret key should allow the legitimate user to correctly use these algorithms.

There exists only one scheme that has been proven perfectly secure: the so-called \gls{OTP}. In that scheme, it is of paramount importance to use a randomly chosen secret key, as long as the plaintext to encrypt, and this key should be used only once. Doing so, a given ciphertext can correspond to any plaintext, and there is no way to decrypt the message without knowing the key. However the \gls{OTP} is not used in practice  due to the complexity of the key transmission and its length. For schemes that cannot be proved perfectly secure, their security level is given by the complexity of the best-known attack against them. For instance, if the best-known attack against an encryption algorithm requires \(2^{80}\) computing cycles, the security level of this scheme is considered to be 80 bits (it's the equivalent of a scheme where the bruteforce of a 80 bit long key is the best attack). The downside of this approach is that it does not account for the memory space complexity: some attacks might perform more efficiently in time than others, but their memory complexity make them completely intractable in practice (\emph{e.g.} enumeration attacks).
In 2022, it is generally admitted that a cryptographic primitive providing 80 bits of security or less can be efficiently broken (within months using modern computers).

Three main primitives exist in modern cryptography : 
\begin{itemize}
    \item Encryption, symmetric or asymmetric ensures confidentiality and authentication
    \item Hash functions ensure integrity
    \item Signature ensures authentication and non-repudiation
\end{itemize}

Combining these primitives allows to ensure every property needed for any specific information. A \gls{HMAC} for instance ensures both authentication and integrity, adding a hash product to encrypted message ensures integrity, etc...

A widely used construction that provides every security property available is the hybrid encryption. It uses an asymmetrical encryption algorithm to agree on a key known only by two entities, providing authentication, then uses that temporary session key to communicate with symmetric encryption. The first step is called an \gls{AKE} or \gls{AKA}, and is the most critical part of the scheme. If it is compromised the entire security provided by the scheme is lost. The transition to symmetrical encryption has two major advantages. Firstly the leak of a session key does not compromise the other sessions, meaning only the leak of a secret key can impact previous and further communications. And secondly, symmetrical encryption is more efficient, quicker, and has less bandwidth overhead. The \gls{TLS} protocol that uses this architecture is the basis of internet security.

\subsubsection{Symmetric encryption}

Symmetric encryption algorithms are based on a shared secret between the communicating parties. When only these two parties know the secret key, and this key has been shared in an authenticated manner, it is considered that the encrypted exchange is authenticated. There exist authenticated symmetrical encryption algorithms that add an authentication mechanism beyond the simple encryption of a message, and \gls{AEAD} that authenticate the non encrypted associated information in a packet.

The current standard for symmetrical encryption is the \gls{AES} algorithm. It won the standardization contest in 2000 but only became widely used after the triple \gls{DES} lost its popularity a few years later. \gls{AES} can be used with 128, 192 or 256-bit keys and encrypts blocks of 128 bits.

\subsubsection{Asymmetric encryption}

Asymmetric encryption uses a pair of keys for each entity. The secret (or private) key is never shared with anyone, while the public key is, as its name suggests, shared with everyone. If someone wants to send a message to an entity, the message is encrypted with the public key of the recipient, and can only be decrypted with the corresponding secret key. This provides confidentiality to the encrypted message, as the secret key is never shared.

Asymmetrical encryption is based on hard mathematical problems, such as the integer factorization for \gls{RSA}~\cite{rsa}, or the discrete logarithm for ElGamal encryption scheme~\cite{elgamal}. Even if the aforementioned problems are not generically proven NP-hard, the best known algorithms solving these problems have at least sub-exponential complexity, making them hard to solve in reasonable time for cryptographic-size parameters.

\subsubsection{Signature and certificates}

Digital signatures are based on asymmetrical algorithms. They are used to validate the authenticity of a message. The signing algorithm takes a message and the secret key, and outputs a signature. To verify it, the signature is put through a verification algorithm with the public key as the other entry. The signature is verified if the output of the verification algorithm corresponds to the original message. This adds authentication and non-repudiation without encrypting the message. This is particularly useful for broadcast (or any message with multiple recipients), where the source of the data only has to sign once and everyone can verify without a sustained communication with the source. In practice, a hash of the message is computed and is the only part signed. This provides integrity and strongly reduces the computation time for the signature and verification processes.

The standard signature schemes used in \gls{TLS} are \gls{RSA-PSS}, \gls{ECDSA}, and \gls{EdDSA}.

Depending on the signature scheme, the size of the keys and the signature varies. The complexity of the signature process and the verification are also different. By choosing the correct algorithm, it is possible to minimize the workload of the signature process or validation (usually at the cost of the other).

To ensure the legitimacy of public keys, a public-key certificate can be used. This is a document attesting the validity of a public key. It includes information about the identity of the owner, as well as the authority that issued the certificate of the key. The authority signs the certificate with its own private key, and the corresponding public key is either directly known by the hardware, or is certified itself by a higher certification authority. This chain will continue up to a certificate authority that is publicly known by the hardware. Currently, the most common certificates are x509. This format can use the three signature schemes allowed by \gls{TLS}.

\subsubsection{Authenticated key exchange/agreement and hybrid encryption}

An \gls{AKE} is based on asymmetric encryption. The public key is available to everyone, and the secret key is known only by its proprietary. A random secret is generated by the entity that wishes to communicate, and is then encrypted with the public key of the recipient. The ciphertext can then be transmitted through an unsecure channel. This process is called a \gls{KEM}. The secret is known only by the two parties because the knowledge of the private key is needed to decrypt the ciphertext and recover the secret. In practice during the \gls{KEM} process, the source also signs a hash of the key to provide mutual authentication, integrity and non-repudiation to the key exchange.

An \gls{AKA} has a similar process, but the secret is not fully generated by one side, it is computed with one's secret key and the other's public key (plus a random number). This process allows for both entities to compute the same secret, which is impossible to obtain without the knowledge of one of the two secret keys. This allows for mutual authentication if the public keys are certified.

The \gls{TLS} protocol uses \gls{DHE}, \gls{ECDHE} and \gls{RSA} (but not in the latest version) for authenticated key exchange.

Once a secret key is shared between two parties, it is derived into a symmetrical session key, used to encrypt communications. This symmetric encryption process is called the \gls{DEM}. The combination of \gls{KEM} and \gls{DEM} is called hybrid encryption, and allows for a secure communication. As symmetric encryption is a lot more efficient than asymmetric encryption, the hybrid encryption scheme optimizes the data expansion and computation power required to check all the security requirements.

\subsubsection{Post-quantum cryptography }

With the development of quantum computing, certain current standards are potentially compromised. For instance, Shor's quantum algorithm~\cite{shor_algorithms_1994} provides a polynomial time solution to the factorization problem, which allows to break any \gls{RSA} key in reasonable time. It can also be used to quantumly compute discrete logarithms in quasi-polynomial time, compromising the security of \gls{DH} like primitives (including those involving elliptic curves). With the simplification of the cryptanalysis of current standards, there is a need for the development of post-quantum solutions.\footnote{Post-quantum algorithms are sometimes referred to as quantum-resistant or quantum-safe in the literature.}

For symmetric algorithms the new possibilities of quantum computing simplify the cryptanalysis but the solutions remain exponential. \gls{AES} in particular is vulnerable to cryptanalysis by Grover's search algorithm~\cite{grover_fast_1996}, which essentially square roots the complexity of an attack (hence halving its security level). Therefore \gls{AES} would only remain relevant for 256-bit keys. The same problem appears for hash functions, where finding collision is made easier with the Grover algorithm. The countermeasure is similar to \gls{AES}, where SHA-256 is potentially vulnerable, SHA-512 will remain secure.

The quantum threat exists mainly for key-exchange and signature schemes. \gls{RSA}-based, \gls{ECC}-based and \gls{DH} schemes will not be secure. The development of quantum-resistant applications has been a field of research since the discovery of the Shor algorithm. There are 5 main (public-key) primitives for post-quantum cryptography:
\begin{itemize}
    \item (Euclidean) Lattices for encryption and signature;
    \item (Error Correcting) Codes for encryption and possibly signature;
    \item Hash functions for signature;
    \item Multivariate polynomials for signature and possibly encryption;
    \item (Elliptic curves) Isogeny for encryption and possibly signature.\footnote{By possibly, the authors mean that there exists propositions in this sense in the literature. However, a substantial fraction of them have proved to be unsecure.}
\end{itemize}

In 2016 the \gls{NIST} has launched a standardization process for post quantum cryptography~\cite{computer_security_division_post-quantum_2017}. The standards have been announced in July 2022. These standards have been thoroughly investigated for vulnerabilities, dismissed or modified during the selection process. However they will take some time to be widely adopted, as the trust in their security by the general public comes with experience.

\subsubsection{Lightweight cryptography }

Each cryptographic scheme has requirements for their usage on hardware. For an encryption algorithm, these requirements are the number of computation cycles and the memory space they use. These numbers are relative to the size of the encrypted message and the level of security. They also depend on the \gls{CPU} architecture (The number of cycles and memory usage varies between an ARM and a x86 \gls{CPU}, or between 32bit and 64bit \gls{CPU}s). The comparison must be made at a similar level of security and on the same hardware to properly evaluate the performance of these algorithms. Characteristics such as ciphertext size overhead compared to the cleartext, and key sizes are also important in the performance evaluation of cryptographic techniques. 

\gls{LWC} represents the schemes that have a smaller computation complexity and/or memory consumption. The goal is to extend the use of cryptography to more constrained devices that are not powerful enough to compute standard algorithms in an acceptable
amount of time.

Lightweight primitives exist for encryption (symmetric and asymmetric), hashes and signatures. Asymmetric encryption and signature schemes can have uneven performance requirements, so they are very lightweight on a constrained side but require a lot more computation on the other side. This is helpful for systems with uneven performance capabilities (\gls{WSN} or \gls{UAS} for example). The powerful central/ground unit of these systems are able to compute more than the outer nodes.

The \gls{NIST} has started a standardization contest for lightweight primitives~\cite{computer_security_division_lightweight_2017}. This process should finish in late 2022 and will define standard lightweight primitives for symmetric and asymmetric encryption, signature and hashes. They will have been thoroughly tested and modified accordingly during the selection process, and therefore will be considered secure enough to be used in a larger scale. This does not mean that they do not have any vulnerability however, and a new powerful zero-day attack could be able to breach the security they provide. But the extensive testing guarantees that no existing method is able to compromise these solutions.

\subsection{Security threats, goals and requirements for \gls{UAS} communication links}
\label{subsec:security_goals}

Information security on \gls{UAS} communication links has several goals. It enables a secured operation and a safe traffic management for \gls{UAV}s. This is the focus of this paper and the properties required for privacy are considered secondary. The protection of mission-specific information is considered to be the responsibility of the operator. The links and their security requirement are presented in Table~\ref{table:links}. The impact column represents the level of importance fro each link. The temporary loss of an important link can be acceptable while any loss of a critical link can impact the safety of the operation and induce physical and material damage. The requirements in parenthesis represent the requirements that do not directly impact operation safety, but are relevant from a privacy and traceability point of view. The security requirements are then discussed and justified in Sections~\ref{subsubsec:uavgcs}, \ref{subsubsec:uavutm} and \ref{subsubsec:uavuav}.

\begin{table}[H]
    \centering
    \begin{tabular}{ | c | c | c | c | } 
      \hline
      Type of & \multirow{2}{*}{Data} & \multirow{2}{*}{Impact} & \multirow{2}{*}{Security Requirements}\\ 
      Communication & & &\\
      \hline
      \multirow{4}{*}{\gls{UAV}-\gls{GCS}} & command & critical & (C), I, A, (NR) \\ 
      \cline{2-4}
      & video & critical & (C), A, I \\ 
      \cline{2-4}
      & telemetry & important & (C), A, I \\
      \cline{2-4}
      & mission data & specific & Non Applicable \\
      \hline
      \multirow{3}{*}{\gls{UTM}-\gls{UAV}} & emergency & critical & I, A, (NR) \\ 
      \cline{2-4}
      & direct & critical & (C), I, A, (NR) \\ 
      \cline{2-4}
      & telemetry & critical & I, A, (NR) \\
      \hline
      \multirow{3}{*}{\gls{UAV}-\gls{UAV}} & relay & important & (C), I, A, (NR) \\ 
      \cline{2-4}
      & routing & important & I, A, (NR) \\
      \cline{2-4}
      & environmental & important & I, A, (NR) \\
      \hline
    \end{tabular}
    \caption{\gls{UAS} links and their security requirements (Confidentiality, Integrity, Authentication, Non-Repudiation).}
    \label{table:links}
\end{table}

\subsubsection{\gls{UAV}-\gls{GCS} link security }
\label{subsubsec:uavgcs}

The link between a \gls{UAV} and its \gls{GCS} is the only link that is technically mandatory to fly. Aside from mission specific data, it can be separated into 3 different links. They are sometimes physically separated, but are usually sent through the same communication mean.

First is the command link that transmits all command inputs or trajectory adjustments from the \gls{GCS} to the \gls{UAV}. The \gls{UAV} can respond with acknowledgements messages. This link is absolutely critical for the safe operation of a \gls{UAS}. If the command link is compromised, the direct control of the \gls{UAV} is in danger. For safe \gls{UTM}, this link has very important needs in terms of authentication and integrity. The \gls{GCS} has to be authenticated as it has overriding authority on the \gls{UAV}s decisions. To avoid being fed false information, the \gls{GCS} needs to trust the \gls{UAV} as well. However, this link needs a sustained availability especially when the \gls{UAV} has no or limited automation. The mission is compromised if the communication with the \gls{GCS} is lost. There are confidentiality needs as well, especially if the the system is organised as the \gls{IoD} architecture presented in Figure~\ref{fig:iod}. An outside observer should not be able to determine which end user controls which \gls{UAV}. The confidentiality is not required to enable a safe \gls{UTM} and is only a privacy concern. 

The video feed is sent from the \gls{UAV} to the \gls{GCS} and is mandatory in any \gls{BVLOS} flight. The pilot cannot control safely the \gls{UAV} without visual input and it is mandatory that a human pilot can take direct control of a \gls{UAV}. The visual information also helps to check the progression of the mission and verify that the \gls{UAV} follows the correct course. Video is also often part of the mission (surveillance, infrastructure monitoring...), but even if mission specific data is left to the discretion of the operator, its critical use for safety makes it a critical link for safe traffic management. This means that authentication and integrity of the video is required. Moreover, video is very sensitive for privacy requirements. The link can contain private information which is why the confidentiality needs are important for privacy.

The \gls{UAV} sends all the telemetry data it collects back to the \gls{GCS}. This link contains a lot of information on the flight, and the data can be used by the \gls{GCS} to modify the flight plan. Compromising this link can indirectly modify the \gls{UAV}'s trajectory. Authentication and integrity are then important to secure the airspace. As a lot of information is transmitted with this link, confidentiality is needed for privacy. The temporary loss of this link is not so critical as it is for the command and video links.

Mission specific data that is not the video used for \gls{UAV} control does not impact the safety of \gls{UTM}. Each operator is responsible for the level of security required for such information. If privacy concerns are raised however, the data will have to remain confidential.

\subsubsection{\gls{UAV}-\gls{UTM} system link security }
\label{subsubsec:uavutm}

\gls{UTM} solutions are still in development, the technical specifications of the links between the \gls{UTM} system and the \gls{UAS} are still not determined. It is known that three types of communications will be required (see Section~\ref{subsec:utm}). The \gls{UTM} system can broadcast emergency messages for real-time airspace organization modifications. It will also be able to contact a specific \gls{UAS} via its \gls{GCS} and eventually the \gls{UAV}. And the \gls{UAV} will need to transmit telemetry information to the \gls{UTM} services.

The emergency broadcast from the \gls{UTM} system is used for last minute modifications in the organization of the airspace. As needed for ensuring the safety of the traffic, new geofences can be enabled, changes of flight authorizations can be put in place. These modifications will impact the operations, forcing the \gls{UAS} to change trajectories, or even grounding the \gls{UAV}s. The availability of this service is critical, as well as authentication, integrity and non-repudiation. Providing confidentiality to this information is not recommended, as it concerns every entity in the airspace, and would need to be interpreted without the security mechanisms.

The direct link between the \gls{UTM} system and a \gls{UAS} is similar to the link an aircraft has with a control tower in aviation. Information concerning only one \gls{UAS} is transmitted. Direct flight recommendations are given by the \gls{UTM} system, for conflict resolution or other traffic management services. Flight plan modifications requests from the \gls{UAS} are also transmitted on this link, as well as the \gls{UTM} system responses. This system will converse with the pilot himself, the \gls{GCS} or the \gls{UAV}'s flight manager (similarly to the different \gls{ATC} communications with an airplane). This system induces course corrections for the \gls{UAV} and is therefore critical to the safety of traffic management. Authentication of both parties, integrity and non-repudiation are needed for securing the traffic, and confidentiality is recommended for privacy issues.

\gls{UAV}s need to be identified remotely, and to communicate part of their telemetry (position, speed...) to the \gls{UTM} system. The \gls{UTM} system needs to be sure it talks to a real \gls{UAV} as its presence can affect the traffic management. For this link, it has been proposed to use a system similar to \gls{ADS-B}, where some telemetry data is openly broadcast without any added security. However commercial aviation has redundancy and does not use this system for the \gls{ATC}~\cite{icao_ads-b_2018}. Other propositions rely on cellular networks to transmit this information. This would improve availability and has a few native security mechanisms. Whatever the communication system used, there is an important need for authentication, integrity and non-repudiation, because vulnerabilities on this link compromise traffic surveillance. The needs of confidentiality are still being debated. As \gls{UAV}s can perform missions where, if the trajectories and the identity of the users were to be leaked, some private information can be extracted (delivery from or to a hospital, transport to sensitive locations...). It is then relevant to propose a security solution that adds confidentiality, or at least some anonymization for the \gls{UAV}-\gls{UTM} link.

\subsubsection{\gls{UAV}-\gls{UAV} link security }
\label{subsubsec:uavuav}

Three types of communications can exist between \gls{UAV}s in \gls{UAS} networks. The network can use \gls{UAV}s as relay nodes for the \gls{UAV}-\gls{GCS} communication or \gls{UTM} services. Organising the network requires routing information to be transmitted. Lastly, \gls{UAV}s can propagate environmental information to neighboring \gls{UAV}s.

When a \gls{UAV} is used as a relay, it becomes a part of the link between the end parties. But to enable safe forwarding of messages, authentication of the relay node is necessary, as well as integrity and non-repudiation. Confidentiality is not as important if the larger link is properly secured, only associated data is left unprotected. For privacy reasons, an operator could want to anonymize the source and destinations of the messages.

In order to be used as relays, the \gls{UAV}s need to use routing protocols to organize dynamically the network. Routing messages are often sent between nodes to update the routes. In order to keep only legitimate nodes in the network, these routing messages need to be authenticated, and integrity and non-repudiation also need to be provided.

Environmental information is used by \gls{UAS} to optimize trajectories. It is important to have it authenticated, as false information can impact the safety of an operation. Integrity and non-repudiation are also important to ensure the legitimacy of these messages. These messages can be broadcast to all neighbouring aircraft, and similarly to emergency messages from the \gls{UTM} system (see Section~\ref{subsubsec:uavutm}), adding confidentiality is not recommended.

\section{\gls{UAS} communication vulnerabilities}
\label{sec:vuln}

\subsection{Sensitive data interception attack}

\subsubsection{Eavesdropping : description and operational impact}

Eavesdropping, sniffing or passive listening is a simple attack that can be performed on any electromagnetic signal. The attacker only has to intercept the communication. In any wireless setting it is achieved simply by setting up a receiver and observing the communication between two parties. If the attacker is then able to extract sensitive information from the signals, the attack is successful. Even when the signal is encrypted, if the identity of the source and destination are not hidden from the attacker, some private information is leaked as well.

Current protocols for telemetry, like the MAVlink are particularly vulnerable to eavesdropping. Kwan \emph{et al.}~\cite{kwon_empirical_2018} detail these flaws in an empirical analysis of the MAVlink protocol. The lack of confidentiality can cause privacy issues, even with the telemetry being the only information leaked. The MAVlink can be transmitted over a secure channel, but as telemetry is not usually considered sensitive, operators do not generally bother and simply use the MAVlink over unencrypted channels.

The video link is particularly vulnerable to data interception, as it can contain sensitive images. The security of the video link varies a lot depending on the \gls{UAV} and its camera. \gls{UAV} manufacturers either include the video feed in their main communication link between the drone and the \gls{GCS} (this is usually done by Wi-Fi enabled \gls{UAV}s), or they use a separate channel to transmit the video (this is common on long range \gls{UAV}s using LoRa or similar protocols for the command link). In this second case, an unprotected method can be used to transmit the video. \gls{NTSC}, \gls{PAL} or \gls{SECAM} are the analog standards for video broadcast used by most CCTVs and off the shelf wireless cameras. \gls{UAV}s that encode their video with those protocols are not protected. Any receiver can decode these signals and therefore no privacy exists with these transmission methods.

\subsubsection{Eavesdropping : mitigation and countermeasures}

Recording and analysing ambient signals is a passive task. It is impossible to detect if someone is listening on a wireless channel. An attacker will always be able to log the signals and try to extract information. The signals can however be protected.

In a multiple \gls{UAV}s system, it is possible to use other nodes to send artificial noise, confusing the eavesdropper. Zhong, Yao and Xu presented a cooperating jamming approach~\cite{zhong_secure_2019} that uses a second \gls{UAV} to mask the confidential information transmitted to the \gls{GCS}. The solution assumes that the system knows the location of the GCS and estimates the eavesdropper's position. It then adjusts the \gls{UAV}'s trajectories overtime to facilitate the jamming. This solution is only viable when the eavesdropper's approximate position is known in advance, and requires the full control of a second \gls{UAV} for the entire communication period.

The physical layer can be protected. Poor and Shaefer~\cite{poor_wireless_2017} present the research in the abilities of the wireless physical layer to provide security. The properties of radio channels such as diffusion and superposition can provide secrecy in wireless data transmission. This is an active field of research, and this solution is not often applied in operational context.

Data confidentiality is added with encryption. If correctly utilized it makes it impossible for the attacker to extract any information from the sniffing. This confidentiality needs to be sustained as the information remains sensitive even long after the flight. The commands and telemetry links can be encrypted using symmetric algorithms. The \gls{AES} algorithm is however not recommended for constrained systems as it induces a lot of computation overhead. Lightweight algorithms have been developed to solve this problem (see Section~\ref{subsec:sym}).

An analog video link cannot be encrypted without being converted into a digital signal. Then encryption becomes possible. As a video feed is a real-time sensitive part of multimedia systems, Chen \emph{et al.} have designed a secure video communication system using chaos based encryption~\cite{chen_design_2018}. This system has excellent real-time performances but is extremely costly in terms of computational power and is therefore not optimized for small \gls{UAV}s. The use of lightweight symmetric algorithms will be more optimized for \gls{UAS}. In order to limit hardware addition, computation overhead and communication overhead, it will be recommended that the video feed from the \gls{UAV} is included in a single link to the \gls{GCS} with the commands.

\subsection{Jamming}

\subsubsection{Communication channel jamming : description and operational impact}

An attacker can interfere with a wireless signal by broadcasting a strong noise over the communication channel. This can cause the disruption of the communication between parties as the signals they want to transmit to the other will be drowned under the noise created by the jammer. The jamming strategy requires the ability of broadcasting signals that will be strong enough to disrupt the real communication upon reception. A jammer that simply broadcasts random noise over a single frequency is called a spot jammer. Jamming simultaneously a large bandwidth by changing frequency extremely rapidly is barrage jamming, but requires an immense power output to still be effective over the whole bandwidth. Some more advanced techniques exist, such as sweep jamming where the jammer jumps from one spot to another in order to disrupt a larger communication channel with similar power output to a spot jammer. The sweep jamming technique does not block the entire signal, and some data will get through. Sweep jammers rely on the fact that they sufficiently disrupt the communication to induce a \gls{DOS}. The jamming strategy works on many different communication links used by \gls{UAV}s. Wi-Fi, LoRa, cellular networks, \gls{GNSS} signals, Bluetooth and many more~\cite{pirayesh_jamming_2022}.

Jamming the different links in a \gls{UAS} causes a \gls{DOS} on the targeted service. This has various consequences depending on the link and the type of \gls{UAS}. For autonomous \gls{UAV}s, the temporary loss of the C2link and video link will compromise any in-flight input by the \gls{GCS}, but the drone will pursue its mission. This causes a threat to flying objects in the vicinity for emergency cases. The detect and avoid systems should mitigate the risk, but a loss of C2link and video link is unacceptable for more than a few seconds for \gls{BVLOS} flights. In the case of less autonomous \gls{UAV}s the loss of the control links is a lot more critical.

Pärlin, Alam and Le Moullec~\cite{parlin_jamming_2018} have developed a protocol-aware jamming system for \gls{UAV}s, improving the service disruption for low jam-to-signal ratios in comparison to sweeping jammers. A simple \gls{SDR} is used to effectively perform a \gls{DOS} attack on the command link of the \gls{UAV}. The authors tested their jammer on two generic remote control systems (FASST and ACCST). By comparing the distance of effectiveness at a fixed power output, they determined that their system was an improvement over the sweeping jammer by a factor of 2.4 for the FASST and 5.7 for ACCST.

\subsubsection{Communication channel jamming : mitigation and countermeasures}

Conventional anti-jamming approaches either mitigate the effects of jamming with limited effectiveness, or rely on detection and deploy a countermeasure while interrupting the communication~\cite{pirayesh_jamming_2022}. 

Mitigation techniques are numerous. The \gls{FHSS} is one of the simplest techniques and is often used in LoRa and Wi-Fi networks, but does not work against smart jammers and is not efficient in terms of spectrum utilization. Specific signal modulations are more resilient to jamming, but also provide a lower bit rate. Research is still ongoing but the solutions like massive \gls{MIMO}-based anti-jamming techniques are too costly to be implemented on a small scale. However coupling a \gls{MIMO} system with the \gls{IRS} technology~\cite{wu_intelligent_2019} would significantly lower the cost of this solution. Results have shown that \gls{IRS}-aided \gls{MIMO} will perform similarly to a massive \gls{MIMO} with a lot less hardware and power usage. \gls{IRS} could then be deployed with smaller \gls{MIMO} systems on much smaller wireless networks, allowing the anti-jamming and the performance improvements offered by massive \gls{MIMO} to be used in smaller \gls{UAS}s.

The jamming detection techniques have evolved from basic noise monitoring to multi-factor learning-based approaches. With factors including the directional parameters of signals the newer detection techniques are able to more effectively detect jammers, and pinpoint their location and characteristics.

The detect and counter strategy is more efficient than mitigation techniques but it results in a temporary loss of the connection. As this is unacceptable for \gls{UAS}s, another approach would be to conceive jamming resilient systems by design. Some physical layer properties could help differentiate real signal from noise, based on directional input and more advanced noise filters.

\subsection{Spoofing attack}

\subsubsection{Communication links spoofing : description and operational impact}

Spoofing is a situation where an attacker successfully usurps the identity of a legitimate user or network node. By taking the identity of a \gls{GCS}, the \gls{UTM} or a \gls{UAV}, an attacker can perform \gls{MITM} attacks, compromising the entire \gls{UAS} through data injection or \gls{DOS}.

Spoofing the command link from the \gls{GCS} to send false information to the drone gives control of the drone to the attacker. In~\cite{westerlund_drone_2019}, this is used on the Parrot AR 2.0 drone. The authors captured the signals used by the \gls{GCS} for different commands and analyzed their structure. They were then able to generate fake commands, spoofing the IP address of the controller and the drone interpreted these messages as legitimate. 

Spoofing the telemetry link from the \gls{UAV} can cause the \gls{GCS} to send course corrections, modifying the trajectory and compromising the \gls{UAV}'s mission.

An attacker can also target the video link, with similar results as the telemetry spoofing. The pilot is influenced to correct the trajectory, compromising the \gls{UAV} and its mission.

The telemetry spoofing attack can also be aimed at the \gls{UTM}. This can simulate fake objects to modify the path of oncoming traffic, create false threats to commercial aviation, disrupting the whole airspace control system.

The \gls{UTM} en-route service itself can also be spoofed. An attacker could disrupt the \gls{UAV}'s trajectory by creating false exclusion zones, removing real ones. It could even send false trajectory modifications to the rest of the \gls{UAS}.

The MAVlink protocol can be particularly vulnerable to spoofing attacks as it does not provide any security. If it is used over a non secure channel it is easily spoofed. The MAVlink can be used for the command link and the telemetry link, meaning that an attacker can use this vulnerability to take full control of the drone.

\subsubsection{Communication links spoofing : mitigation and countermeasures}

A smart \gls{IDS} can detect spoofing~\cite{anthi_supervised_2019}. It requires monitoring from a central overseeing position to be most effective. This system works better on Ad-Hoc networks, and is based on the identification of malicious packets. \gls{ML} techniques can then be used to distinguish legitimate nodes from spoofed ones. New spoofing techniques will however potentially work undetected.

Spoofing is strongly countered by proper authentication. The use of cryptographic primitives with correct implementation ensures the authentication of all parties. Over the internet the \gls{TLS} protocol ensures such properties with the use of https. The particular conditions of \gls{UAS} makes it incompatible with \gls{TLS} due to the bandwidth overhead and computational power required for a sustained used of \gls{TLS}. Lightweight authentication techniques are detailed in Section~\ref{subsec:auth}.

\subsection{False information dissemination (\gls{UAV} to \gls{UAV})}

\subsubsection{False data injection : description and operational impact}

False data injection is an attack in which false information is transmitted from a compromised \gls{UAV}. This can concern false environmental conditions forcing legitimate \gls{UAV}s to modify their trajectories, reducing the success rate of their missions as well as their survivability. This can also concern the \gls{ADS-B} in the cases where it is used to avoid collisions. Since \gls{ADS-B} is an open broadcast with no authentication protocols, an attacker can create a fake object at critical positions or broadcast counterfeit positions for the legitimate \gls{UAV}s by impersonating them.

\subsubsection{False data injection : mitigation and countermeasures}

The \gls{IDS} proposed in~\cite{sedjelmaci_hierarchical_2018} is able to detect \gls{ADS-B} attacks by cross-checking the \gls{ADS-B} information with the location estimated with signal intensity and round trip time data. This \gls{IDS} also verifies the coherence of environmental information with the other drones in the network, but this depends on the sufficient presence of legitimate users in the area. The \gls{IDS} can raise an alarm and the \gls{UAS} will deal with the compromised node.

These attacks can be countered by cryptographic methods (see Section~\ref{sec:crypto}) that ensure authentication and integrity. If the \gls{UAV}s are trustworthy and the integrity of the signals is verified, the attack is no longer possible as an outside attacker will not be able to counterfeit authenticated data. 

\subsection{Malicious detection and identification}

\subsubsection{Malicious detection : description and operational impact}

With the arrival of ADS-B, Remote-ID and other \gls{UAV} identification frameworks, some privacy concerns have been raised as \gls{UAV} detection is easily performed. If \gls{UAV}s and their mission can be easily identified, attackers can detect and physically threaten relevant \gls{UAV}s. This topic has been debated, with the \gls{NBAA} raising concerns about the privacy issues created by the mandatory use of Remote-ID in the United States~\cite{noauthor_nbaa_2021}.

Even without such systems, works such as~\cite{ezuma_micro-uav_2019} allow for remote detection and identification of \gls{UAV}s based on \gls{RF} fingerprints in their controller's signals. The authors used \gls{ML} techniques to classify the signals and identify 14 types of \gls{UAV}s. In~\cite{ezuma_detection_2020}, the same authors propose a scheme where these identifications are made possible with interference.

Shoufan \emph{et al.} have proposed a system that is able to identify a \gls{UAV}'s pilot based on flight behaviour~\cite{shoufan_drone_2018}. Using \gls{ML} techniques they are able to identify 20 pilots with 90\% accuracy. This is only possible when the pilot is directly commanding the drone, and requires a dataset for all pilots to be effective. 

\subsubsection{Malicious detection : mitigation and countermeasures}

The authentication scheme behind Remote ID and other potential identification frameworks can be protected for privacy. In \cite{alsoliman_privacy-preserving_2020}, Alsoliman, Rabiah and Levorato developed an extension of the Remote ID framework with privacy-preserving capabilities. This solution allows for an anonymized authentication where the pilot and the operator's identities are known only by the authorities. It is also able to verify the flight permissions of a \gls{UAV} flying in a specific area without revealing the \gls{UAV}'s entire flight path. In this work, the flight plans are sliced, and the Remote-ID messages are modified with a pseudonymous certificate, allowing for \gls{UAV} authentication and authenticated flight permission for the current area.

The \gls{FAA} proposed the Privacy \gls{ICAO} Address~\cite{noauthor_ads-b_nodate} to respond to the privacy concerns raised by \gls{UAV} operators regarding the use of \gls{ADS-B}. This solution provides operators with temporary \gls{ICAO} addresses for their aircraft which are not assigned to the owner in the civil aviation registry. This enables owners to broadcast \gls{ADS-B} messages with anonymized identifiers.

\subsection{Routing attacks}

\subsubsection{Black hole and gray hole : description and operational impact}

The black hole attack is a network attack were a compromised node presents itself as the route for all destinations, but then deletes all the packets it receives, inducing a \gls{DOS} on the network. This is done by replying favorably to any route request message received, hence attracting all future packets to the compromised node.

The gray hole attack is an improved version of the black hole, were in order to not be detected and excluded from the network, the compromised node will only drop the packets with a certain probability. Alternatively gray hole attacks can be selective with the dropping of packets. This results in an increase in communication delay, or to a full \gls{DOS} depending on the network characteristics and the importance of the data transmitted.

\glspl{VANET} are particularly vulnerable to these attacks. The dynamic nature of the network topology causes routing tables to be continuously updated, thus increasing the potential impact of attacks using routing vulnerabilities.

\subsubsection{Black hole and gray hole : mitigation and countermeasures}

In~\cite{sedjelmaci_hierarchical_2018}, the IDS is able to detect black and gray hole attacks thanks to a trusted supervisor that monitors packet loss. The system raises an alarm when a node is failing to forward a number of packets superior to a threshold representing natural packet loss.

Applying cryptographic primitives to routing messages is an alternative if the compromised node is not an authenticated part of the network~\cite{ponsam_survey_2014}. This is also a mechanism that protects against black hole attacks where the attacker spoofs a legitimate node of the network. The authentication mechanism is however vulnerable to \gls{DOS}, as the verification delay can be used to slow the nodes by flooding them with packets. This effect is mitigated by the use of lightweight cryptography (see Section~\ref{subsec:auth}).

\subsection{Wi-Fi specific attacks}

\subsubsection{Wi-Fi enabled attacks against \gls{UAS} : description and operational impact}

Wi-Fi enabled \gls{UAS}s are mostly recreational and usually have little to no focus on information security. Their security properties rely mostly on the security properties of the communication link, the Wi-Fi. Some do not have any security and have an open \gls{AP} that is easily hijacked. Others use old protocols such as WEP that have been rendered obsolete by modern cryptanalysis. Recent Wi-Fi enable \gls{UAV}s generally use a private \gls{AP} secured by WPA2, but they are still extremely vulnerable. The use of WPA3 is not yet widespread, but is more strongly defended with the correct settings. WPA2 is vulnerable, particularly when the credentials are not randomized. Some bruteforce attacks on the secret key such as the PMKID attack will even work at low resource with weak passwords. Rogue \gls{AP} can be used for \gls{MITM} attacks.

He, Chan and Guizani~\cite{he_drone-assisted_2017} have performed standard Wi-Fi attacks on \gls{UAS}. The WEP protocol is extremely weak, only the WPA2 can remain strong against low resource bruteforce attacks. They used the Aircrack-ng suite to compromise the Wi-Fi network used by the drone. It is also possible to crack the key using ARP replay attacks, as the drone will always respond to ARP requests with variable Initialization Vectors (IVs), which helps in cracking the key. A deauthentication attack is also effective, the attacker can send deauthentication frames, disrupting the connection with the legitimate user.

Westerlund and Asif~\cite{westerlund_drone_2019} evaluated two commercial Wi-Fi enabled \gls{UAV}s : the Parrot AR 2.0 and the Cheerson CX-10W. They investigated several Wi-Fi attacks to test the security of the communication links of these drones. In particular they performed \gls{DOS} attacks by deauthentication, \gls{MITM} by deauthentication and reconnection to fake \gls{AP}, deauthentication and reconnection to an illegitimate controller. Most of these attacks are induced by the lack of pilot credential in the native system. The authors then built a \gls{UAV} disabler for the Parrot with a Raspberry Pi3, using their previous findings to exploit the communication vulnerabilities.

\subsubsection{Wi-Fi enabled attacks against \gls{UAS} : mitigation and countermeasures}

Wi-Fi enabled drones can be protected by the Wi-Fi security features. Using proper security provided by the WPA2 protocol, with a strong random key mostly protects against low resource attacks. 

In~\cite{hooper_securing_2016}, Hooper \emph{et al.} explore the security flaws of the Parrot Bebop \gls{UAV} and propose a security framework to defend the \gls{AP} of Wi-Fi enabled \gls{UAV}s. This research proposed solutions against the ARP poisoning attack, as well as \gls{DOS} created by the presence of multiple potential controllers. The authors discuss the importance of a multi-layered security framework against zero-day vulnerabilities. They argue that the low importance given to security aspects of small \gls{UAS} causes a lot of zero-day vulnerabilities to exist, and that only a multi-layer security approach remains protected when a new attack is discovered.

\subsection{Port scanning and open services}

\subsubsection{Open ports and services : description and operational impact}

In most off-the-shelf \gls{UAV}s, once connected to the wireless network, the services used to control the drone, receive video or other data, or transmitting software updates run on different ports. The nmap tool can scan those ports, and determine the ones that are used and open for communication. The telnet and \gls{FTP} ports are often opened and, by design, unprotected. Rudo and Keng~\cite{rudo_consumer_2020} have performed these tests on a commercial recreational \gls{UAV}, the Parrot Bebop 2. They found that the \gls{FTP} ports (20 and 21) were used by the system and performed a fuzzing attack onto the FTP control port (21). This attack relies on randomly sending protocol-specific keywords until getting a response from the service running on the targeted port.

In~\cite{westerlund_drone_2019}, the open telnet port allows direct root access on the Parrot AR drone 2.0, without credentials. The \gls{FTP} service is also enabled, and allows file transfer during flight simply by knowing the drone's static IP.

\subsubsection{Open ports and services : mitigation and countermeasures}

The simple application of good practices resolves most of the problems. Simply adding a credentials system and being preventive by shutting down services during flight greatly reduces the risks presented above. In the cases of~\cite{rudo_consumer_2020} and~\cite{westerlund_drone_2019}, \gls{FTP} service is only used for software updates, and should not be running in flight mode. If ssh, telnet, ftp or other services are needed for the \gls{UAS}'s operation, manufacturers should not allow their access with no credentials as it offers easy hijacking possibilities to illegitimate users.

\section{Securing \gls{UTM} two-way communications with encryption}
\label{sec:crypto}

\subsection{Description}

Cryptography can provide confidentiality, integrity, authentication and non-repudiation to information. The use of encryption, signature and hash functions allows for different properties to be added on a message. For \gls{UAS} communications, those properties are needed throughout the different channels. Authentication needs asymmetrical algorithms in a complex network, as the quadratic number of keys required for symmetrical encryption or password authentication is not acceptable. Each member has a secret key used to generate temporary session keys with another member, which are then used for symmetric encryption. This process called \gls{AKA} or \gls{AKE} is the basis for authentication, integrity, confidentiality and non-repudiation. This method is optimal for security, but requires a two-way communication. When only integrity is required, a hash function is used. This solution is valid only when the only threat is caused by involuntary phenomenons (such as natural transmission loss) and not purposeful manipulation. For integrity and authentication without encryption, a signature scheme is preferred. This is needed when a system does not require confidentiality. For broadcast messages, where receivers are given the choice not to use the security mechanisms for performance improvements or any other reason, this solution is highly preferable.

\subsection{Providing authentication to \gls{UAS} actors with encryption}
\label{subsec:auth}

\subsubsection{Description}

Due to the poor performance encountered in \gls{UAV} networks, classic authentication techniques are not preferred. As an example, the \gls{TLS} handshake has too many steps, and uses RSA or \gls{DH} asymmetrical algorithms to exchange the session keys. These primitives are not efficient in terms of computation and are vulnerable to future quantum threat. Many lightweight \gls{AKA} schemes have been proposed in the \gls{UAV} field as well as other fields with similar constrains.

\subsubsection{\gls{UAS} specific research}
\label{subsubsec:uas}

A lot of research has focused on improving the performance of cryptographic solutions for authentication in a \gls{UAV} context. Lightweight \gls{AKA}s have been proposed by several authors. A lot of them use the \gls{IoD} control architecture as a model.

Wazid \emph{et al.} have proposed an \gls{AKA} scheme for the \gls{IoD} architecture~\cite{wazid_design_2019}. It provides a mutual authentication between a drone and a user. The user authentication is done with three factors : user password, smart card (replaced by a mobile device) and user biometrics. Registration needs to be done over a secure channel, as the server generates secrets for users and drones. The scheme is lightweight, and uses only hash functions and \gls{XOR} operations (excluding the biometrics extraction). The server generates and stores every secret key, and is a \gls{SPOF} of the scheme. User password and biometrics can be updated, and drone-to-drone session keys can be generated. This scheme was tested with the AVISPA tool, and compared with other lightweight \gls{AKA}s through simulation, using SHA-1 for the hash function in their comparison. As this protocol is designed for an \gls{IoD} architecture, the hardware is owned by a service provider who has overarching capabilities on the drones through the server that manages all secret keys.

In~\cite{srinivas_tcalas_2019}, Srinivas \emph{et al.} have proposed TCALAS, a lightweight \gls{AKE} for the \gls{IoD} architecture. It only uses hash primitives, and \gls{XOR} operations. The ground station server manages every key for drones and users. This scheme provides mutual authentication, parameter updates and user or drone revocation. A formal security analysis shows the resistance against various attacks. Performance comparison have been done with the SHA-1 algorithm as the hash function.

Ali \emph{et al.} have proposed an improvement over TCALAS in~\cite{ali_securing_2020} called iTCALAS. It is scalable, allows for multiple flying zones, provides user anonymity and defends against stolen verifier attacks. The security is evaluated with a formal security analysis, and with the ProVerif tool.

The Lake-IoD \gls{AKE} is a solution developed by Tanveer \emph{et al.} for the \gls{IoD} architecture~\cite{tanveer_lake-iod_2020}. It is a 3-party \gls{AKE} protocol based on AEGIS (see Section~\ref{subsec:sym}) and SHA-256 algorithms. User parameters can be updated, drones and users can be dynamically deployed and revoked by the system. This scheme requires the management system to store keys for every user and \gls{UAV}. The management system is always involved in communications and is a \gls{SPOF}.

There have been propositions using internal physical properties of the hardware to generate unclonable reproducible randomness. This allows any integrated circuit to provide a physically defined footprint that cannot be reproduced. This is called a \gls{PUF}, and has been used as a physical authentication basis of \gls{UAV}s by Pu and Li \cite{pu_lightweight_2020}. They use a \gls{PUF} of the drone to create a challenge response pair that is stored only in the \gls{GCS}. This challenge is only reproducible by the original hardware and a new one is generated each time an authentication mechanism is completed.

In~\cite{zhang_lightweight_2020}, Zhang \emph{et al.} propose an \gls{AKA} for the \gls{IoD} based on passwords. It requires a hash primitive and bitwise \gls{XOR} operations and is faster and lighter than the schemes it compares to.

Cho \emph{et al.} proposed an authentication framework called SENTINEL, based on \gls{HMAC} that was 3 times faster than the equivalent \gls{TLS} solution~\cite{cho_sentinel_2020}. However in~\cite{jan_design_2021}, it is proven vulnerable against several attacks. In this paper, Jan \emph{et al.} also remark that the SENTINEL protocol lacks dynamic node addition and revocation. \cite{jan_design_2021} proposes an improvement that is then formally analysed by ProVerif 2.02.

Nikooghadam \emph{et al.} propose a provably secure lightweight authentication scheme for the \gls{IoD} architecture~\cite{nikooghadam_provably_2021}. It is based on \gls{ECC} and hash primitives. A formal security analysis is performed with the Scyther tool. The scheme's performance is compared when implemented with SHA-1 and 160-bit elliptic curve.

\subsubsection{\gls{IoT} research}

In 2014, Turkanivic, Brumen and Höbl have proposed a lightweight authentication scheme based on hash functions and \gls{XOR} operations~\cite{turkanovic_novel_2014}. It provides mutual authentication between parties, password protection, free password choice, password changing and dynamic node addition. It is a two-factor (user password and smart card) authentication scheme. Unfortunately, it has since been proven to be vulnerable to stolen smart card and \gls{MITM} attacks~\cite{farash_efficient_2016}. It also does not provide properties such as forward and backwards secrecy and untraceability. \cite{farash_efficient_2016} tries to correct these issues at the cost of some performance. Amin \emph{et al.}~\cite{amin_design_2016} proved that some vulnerabilities remained and corrected them.

In~\cite{suarez-albela_practical_2018}, Suarez-Albela \emph{et al.} made a performance comparison between \gls{ECC} and \gls{RSA} for signature algorithms. It shows that \gls{ECDSA} is two times faster for low security (128 bits) and that the gap grows with the level of security.

In~\cite{roy_chaotic_2018}, Roy \emph{et al.} propose an anonymous three-factor user authentication scheme. It is based on a hash primitive, a fuzzy extractor and Chebyshev maps. Chebyshev chaotic maps are  better suited to resource-constrained devices than \gls{ECC} and \gls{RSA}. A formal security analysis of this scheme has been done with the ProVerif 1.93 tool, and its performance is evaluated against other chaotic map-based solutions.

\subsubsection{Medical field research}

The medical field has similar constraints to \gls{UAS}'s for certain systems. For example, healthcare sensors or pacemakers are critically important and resource limited. These characteristics allow the protocols developed for these systems to also be relevant for \gls{UAS} security. 

In \cite{challa_efficient_2018}, an \gls{ECC} based \gls{AKE} is proposed for wireless healthcare sensors. This protocol provides mutual authentication, and a three-factor authentication for users : smart card, password and biometrics. It supports smart card revocation and parameters update methods. three-factor authentication protocols have been proposed for \gls{UAS}s, by swapping the smart card for a mobile device (see Section~\ref{subsubsec:uas}). The lightweight properties of the system have been optimized, and the computational and bandwidth overheads are limited.

\subsubsection{Ad-Hoc networks research}

\gls{MANET}, \gls{VANET} in particular, are difficult to secure with little computational power and bandwidth. Thus some research aims at developing solid authentications schemes for them.

He \emph{et al.} have developed a privacy preserving authentication scheme providing mutual authentication~\cite{he_efficient_2015}. Due to the nature of \gls{VANET}, the scheme aims at being as lightweight as possible. This scheme uses \gls{ECC} and hash primitives, and is suitable for \gls{VANET} deployment. The implementation uses SHA-1 and 160-bit elliptic curve.

\subsubsection{\gls{AKE}/\gls{AKA} performance comparison}

To compare the different \gls{AKA}'s performance, there are different metrics to take into account :
\begin{itemize}
    \item The properties like mutual authentication, quantum resistance, multiple factors, perfect forward secrecy, secret key storage and other implementation specifics will determine the attacks that the protocols remain vulnerable to.
    \item The level of security is determined as the complexity of the best known attack against the system (for example, attacking a 1024 bits \gls{RSA} public key has a complexity of roughly \(2^{80}\), yielding 80 bits of security).
    \item The bandwidth overhead is determined by the number of supplementary bytes sent during the authentication phase (not during registration/initialization).
    \item The computation overhead is usually a time comparison between standards and new protocols. It may be less representative as it is often performed on different CPUs. The relative overhead compared to the standards are used in this comparison even if it is not accurate.
    \item The memory overhead is the size taken in memory by the protocol parameters before and during authentication (IVs, keys...). This variable depends on the scale used for the results provided and is probably the least homogeneous metric.
\end{itemize}

Security comparison : Table~\ref{table:security} compares the security properties of the different authentication protocols presented above.
\checkmark means the protocol provides the feature or protects against the designed attack. An empty cell means the feature is not supported or that the protocol is vulnerable to the attack.

\begin{table}[H]
    \begin{adjustwidth}[]{-8cm}{-8cm}
        \centering
        \begin{tabular}{ | c | c c c c c c c c c c c c c c c c | } 
          \hline
          & \rotatebox{90}{Mutual authentication} & \rotatebox{90}{Public-Key authentication} & \rotatebox{90}{Two-factor authentication} & \rotatebox{90}{Three-factor authentication} & \rotatebox{90}{Anonymity} & \rotatebox{90}{Traceability attack} & \rotatebox{90}{Quantum resistance} & \rotatebox{90}{Revocability} & \rotatebox{90}{Formal security analysis} & \rotatebox{90}{Credential updates} & \rotatebox{90}{\gls{MITM}} & \rotatebox{90}{Replay attack} & \rotatebox{90}{Impersonation} & \rotatebox{90}{Password guessing} & \rotatebox{90}{Privileged insider} & \rotatebox{90}{Stolen authentication factor }  \\ 
          \hline
          \cite{wazid_design_2019} & \checkmark & & \checkmark & \checkmark & & \checkmark & & & \checkmark & \checkmark & \checkmark & & & \checkmark & & \checkmark \\
          \hline
          \cite{srinivas_tcalas_2019} & \checkmark & & \checkmark & \checkmark & & & & \checkmark & \checkmark & \checkmark & \checkmark & \checkmark & \checkmark & \checkmark & \checkmark & \checkmark \\
          \hline
          \cite{ali_securing_2020} & \checkmark & & \checkmark & \checkmark & \checkmark & \checkmark & & \checkmark & \checkmark & \checkmark & \checkmark & \checkmark & \checkmark & \checkmark & \checkmark & \checkmark \\
          \hline
          \cite{tanveer_lake-iod_2020} & \checkmark & & \checkmark & \checkmark & \checkmark & \checkmark & \checkmark & \checkmark & \checkmark & \checkmark & \checkmark & \checkmark & \checkmark & \checkmark & \checkmark & \checkmark \\
          \hline
          \cite{pu_lightweight_2020} & \checkmark & & & & & & \checkmark & \checkmark & & \checkmark & \checkmark & \checkmark & \checkmark & \checkmark & & \checkmark \\
          \hline
          \cite{zhang_lightweight_2020} & \checkmark & & \checkmark & & \checkmark & \checkmark & \checkmark & & & & \checkmark & \checkmark & \checkmark & \checkmark & \checkmark & \checkmark \\
          \hline
          \cite{cho_sentinel_2020} & \checkmark & & \checkmark & \checkmark & \checkmark & \checkmark & \checkmark & \checkmark & \checkmark & & \checkmark & \checkmark & \checkmark & \checkmark & & \checkmark \\
          \hline
          \cite{jan_design_2021} & \checkmark & & & & \checkmark & \checkmark & \checkmark & \checkmark & \checkmark & \checkmark & \checkmark & \checkmark & \checkmark & \checkmark & \checkmark & \checkmark \\
          \hline
          \cite{nikooghadam_provably_2021} & \checkmark & \checkmark & & & \checkmark & \checkmark & & & \checkmark & \checkmark & \checkmark & \checkmark & \checkmark & \checkmark & \checkmark & \checkmark \\
          \hline
          \cite{turkanovic_novel_2014} & \checkmark & & \checkmark & & & & \checkmark & & \checkmark & & & & & & & \\
          \hline
          \cite{farash_efficient_2016} & \checkmark & & \checkmark & & & & \checkmark & & \checkmark & & \checkmark & \checkmark & \checkmark & \checkmark & \checkmark & \checkmark \\
          \hline
          \cite{amin_design_2016} & \checkmark & & \checkmark & \checkmark & \checkmark & \checkmark & \checkmark & \checkmark & & \checkmark & \checkmark & \checkmark & \checkmark & \checkmark & \checkmark & \checkmark \\
          \hline
          \cite{roy_chaotic_2018} & \checkmark & \checkmark & \checkmark & \checkmark & \checkmark & \checkmark & & & \checkmark & \checkmark & \checkmark & \checkmark & \checkmark & \checkmark & & \checkmark \\
          \hline
          \cite{challa_efficient_2018} & \checkmark & \checkmark & \checkmark & \checkmark & \checkmark & \checkmark & & \checkmark & \checkmark & \checkmark & \checkmark & \checkmark & \checkmark & \checkmark & \checkmark & \checkmark \\
          \hline
          \cite{he_efficient_2015} & \checkmark & \checkmark & & & \checkmark & \checkmark & & \checkmark & \checkmark & \checkmark & \checkmark & \checkmark & \checkmark & \checkmark & \checkmark & \checkmark \\
          \hline
        \end{tabular}
        \caption{Security features of presented protocols.}
        \label{table:security}
    \end{adjustwidth}
\end{table}

Table~\ref{table:operations} presents the relative cost of computation for different operations at a similar security level (80 bits) as defined by~\cite{srinivas_tcalas_2019},~\cite{he_efficient_2015} (for the \gls{ECC} point addition only) and~\cite{roy_chaotic_2018} (for Chebyshev Map operation). These values are normalized with a hash computation considered to cost 1 T (where T the time complexity of a hash function) and the other values pondered and rounded to the closest integer. In this context, the cost of a operation costing an amount inferior by several orders of magnitude to the hash function is approximated to 0 T. This is the case for elementary operations : 

\begin{table}[H]
    \centering
    \begin{tabular}{ | c | c | } 
      \hline
      Operation & Computation cost\\ 
      \hline
      Hash function ($T_H$) & 1 T \\ 
      \hline
      Elementary operations (\gls{XOR}, addition...) & 0 T \\ 
      \hline
      \gls{ECC} point addition ($T_{ECC_{PA}}$) & 18 T \\
      \hline
      \gls{ECC} multiplication ($T_{ECC_M}$) & 53 T \\
      \hline
      Fuzzy Extraction ($T_{FE}$) & 53 T \\
      \hline
      Symmetric encryption/decryption ($T_{sym}$) & 2 T \\
      \hline
      Chebyshev Map Operation ($T_{Cheb}$) & 17 T \\
      \hline
    \end{tabular}
    \caption{Computational cost of basic operations.}
    \label{table:operations}
\end{table}

\begin{table}[H]
    \centering
    \begin{tabular}{ | c | c | } 
      \hline
      Abbreviation & Meaning \\ 
      \hline
      t & size of timestamp \\ 
      \hline
      h & size of hash function output \\ 
      \hline
      ec & size of elliptic curve input \\ 
      \hline
      id & size of node identity \\
      \hline
      crp & size of challenge response input \\
      \hline
      cheb & size of Chebyshev polynomial input/output \\
      \hline
    \end{tabular}
    \caption{Abbreviations for the bandwidth overhead.}
    \label{table:bandwidth}
\end{table}

Performance comparison : Table~\ref{table:authperfo} presents the performance of the authentication protocols presented above. When possible, the values in this table are independent from the variables depending on the algorithms chosen by the implementation. If only numerical values are provided, a similar security level is considered. The table contains a relative computational performance comparison with the values defined in Table~\ref{table:operations}.

The values followed by * come from comparisons done outside of the original paper, by different authors. The storage overhead is often undisclosed (ND).

\begin{table}[H]
    \begin{adjustwidth}[]{-8cm}{-8cm}
    \centering
    \footnotesize
        \begin{tabular}{ | c | c | c | c | c | } 
          \hline
          & Computation Overhead Analytic  & Total Bandwidth  & Number of & Storage overhead  \\
          & and Evaluated (drone side) & Overhead & Messages & (drone side) \\
          \hline
          \cite{wazid_design_2019} & 31$T_H$ + $T_{FE}$ = 84 T (7$T_H$ = 7 T) & 10h + 3t & 3 & \(>\)2888b (\(>\)480 b) \\
          \hline
          \cite{srinivas_tcalas_2019} & 30$T_H$ + $T_{FE}$ = 83 T (7$T_H$ = 7 T) & 9h + 3t & 3 & 2888b* (640b*) \\
          \hline
          \cite{ali_securing_2020} & 24$T_H$ + $T_{FE}$ + 3$T_{sym}$ = 83 T (7$T_H$ = 7 T) & 10h + 3t & 3 & 2888b* (640b*)\\
          \hline
          \cite{tanveer_lake-iod_2020} & 11$T_H$ + $T_{FE}$ + 6$T_{sym}$ = 76 T (3$T_H$ + 2$T_{sym}$ = 7 T) & 1376b (fixed) & 3 & 2088b (512b) \\
          \hline
          \cite{pu_lightweight_2020} & approx. 1/10 of \cite{wazid_design_2019} = 9 T (approx 1/3 of \cite{wazid_design_2019} = 3 T) & 2h + 3id + 4crp & 3 & ND \\
          \hline
          \cite{zhang_lightweight_2020} & 24$T_H$ = 24 T (7$T_H$ = 7 T) & 9h + t & 3 & 2756b* \\
          \hline
          \cite{cho_sentinel_2020} & approx. 3 of \cite{wazid_design_2019}* (according to \cite{jan_design_2021}) = 252 T* (21 T*) & 4944b & 3 & 2488b \\
          \hline
          \cite{jan_design_2021} & 13$T_H$ + 4$T_{sym}$ = 21 T (7$T_H$ + 2$T_{sym}$ = 11 T) & 3t + 12h & 4 & 1120b \\
          \hline
          \cite{nikooghadam_provably_2021} & 19$T_H$ + 4$T_{ECC_M}$ = 231 T (5$T_H$ + 2$T_{ECC_M}$ = 111 T) & 8h + 3t + 6ec & 3 & ND \\
          \hline
          \cite{turkanovic_novel_2014} & 19$T_H$ = 19 T (5$T_H$ = 5 T) & 15h + 10t & 4 & 4336b*\\
          \hline
          \cite{farash_efficient_2016} & 32$T_H$ = 32 T (7$T_H$ = 7 T) & 16h + 6t & 4 & ND (2128b) \\
          \hline
          \cite{amin_design_2016} &  32$T_H$ = 32 T (7$T_H$ = 7 T) & 15h + 3t & 6 & ND \\
          \hline
          \cite{roy_chaotic_2018} & 14$T_H$ + $T_{FE}$ + 3$T_{Cheb}$ = 115 T (5$T_H$ + $T_{Cheb}$ = 22 T) & 3h + 2t + 3cheb & 2 & ND \\
          \hline
          \cite{challa_efficient_2018} & 19$T_H$ + $T_{FE}$ + 3$T_{ECC_M}$ = 115 T (5$T_H$ = 5 T) & 4h + 4t + 3ec & 3 & 1776b* (320b*) \\
          \hline
          \cite{he_efficient_2015} & 10$T_H$ + 4$T_{ECC_{PA}}$ + 12$T_{ECC_M}$ = 718 T (5$T_H$ + 2$T_{ECC_{PA}}$ + 6$T_{ECC_M}$ = 359 T) & 2h + 2t + 12ec & 2 & ND \\ 
          \hline
        \end{tabular}
    \end{adjustwidth}
    \caption{Performance comparison of presented protocols.}
    \label{table:authperfo}
\end{table}

Commonly, the implementation of these solutions was done with the minimal level of security of 80 bits. This is not acceptable for sensitive information. The analytical comparison is more relevant for evaluating a scheme's efficiency, as it does not depend on the level of security. In~\cite{cho_sentinel_2020} and~\cite{tanveer_lake-iod_2020} however, the algorithms are inherently linked to the architecture, and a lot more work is required to change the level of security. For the memory overhead, the details are not provided, thus making the results fixed to the level of security used in the implementation of each solution. Unfortunately, the use of the SHA-1 algorithm as a collision resistant hash function is wrong. SHA-1 is flawed, and has been proven non collision-resistant in 2004, and even more so with chosen-prefix attacks by Leurent and Peyrin in~\cite{leurent_collisions_2019}. Moreover, 80 bits is not enough to counter bruteforce attacks with lots of resources. The increase of security level does not change the analytical results, but changes the memory storage given by experimental approach.

These performance comparisons are not absolute. Given dedicated hardware, the computational overhead of certain operations can change relatively to the others. This is only a rough estimate of the performance offered by these solutions compared to one another. It is also important to note that while the schemes using authentication by password seem to offer more security features and are more performant, they do present a few drawbacks. Firstly the addition of new elements requires an initial authentication that is not considered in this comparison. Moreover, communication between two nodes cannot be initiated without contacting the management station, as it is the only element able to authenticate every single node. Should the passwords leak from the management server, an attacker will be able to take full control of the system. With public key infrastructures these problems do not exist, as private keys are known only by their proprietary and only approved by the management system. These schemes are more expensive in terms of performance, but when coupled with a certification system, are not so dependant on the management server. Provided security features also seem to be inferior to the password authentication schemes, but most are achievable at the cost of more memory, computation and bandwidth. Increasing the scale is also a lot easier than with password authentication. It does not require a massive increase in the management server's capabilities. Authentication with external elements like \gls{UTM} systems is more reliable. Indeed, it offers the possibility of contacting nodes individually without passing through the management server for every initial communication. The other major problem aside from the cost is their lack of quantum resistance, which is mainly due to the recency of the development of quantum resistant asymmetrical solutions. 

Password/hash based authentication protocols will work best for the control of a fleet belonging to a single operator, but is not adapted to communication with external entities, and thus cannot fulfill the requirements for data links directed to other drones, other aircraft and the \gls{UTM} systems. It is not acceptable to have to contact the management station in order to establish a time-sensitive communication necessary for safety.

The research works presented above will not all be compliant with the security requirements of large scale \gls{UTM} infrastructures. They have vulnerabilities that are not acceptable, a lack of scalability, or a lack of flexibility. But they can all be relevant for different infrastructures. The password authentication schemes are very good in terms of performance, and for nano \gls{UAV} swarms, they can be preferred over anything else. For larger \gls{UAV}s in reduced density, the performance constrains are not so strict, and it might be preferable to use the reliable non lightweight solutions. However with the increase of traffic density, the bandwidth available for communications will be forcefully limited regardless of the computational and memory abilities of the aircraft. Focusing on bandwidth overhead reduction is the most important objective for enabling a safe \gls{UTM} in traffic heavy areas.

\subsection{Symmetric encryption lightweight solutions}
\label{subsec:sym}

\subsubsection{Description}

In a secure communication system, most of the data is transmitted through symmetric encryption, using session keys obtained during the authentication. This section presents relevant lightweight alternatives to \gls{AES} with their advantages and drawbacks.

\subsubsection{Usual algorithms}

The Chacha20 algorithm is the preferred current lightweight symmetric encryption algorithm for \gls{IoT}~\cite{kane_security_2020}. It performs better than the \gls{AES} standard in terms of energy cost, time cost and memory usage. Chacha20 is 2 to 3 times more efficient than any version of \gls{AES}-128 for time and energy consumption, as well as using less memory for a shorter time. It uses 256 bits keys (or 128 bits keys that are expanded to 256) and are twice as secure as \gls{AES}-128. However with the use of hardware acceleration, the computation time of \gls{AES} can be greatly improved.

An improvement over the Chacha20 algorithm has been proposed~\cite{mahdi_improved_2021} where the complexity of the bruteforce attack grows from \(2^{248}\) to \(2^{512}\), while only reducing the performance by about 20 \%. 512 bits of security is however not needed in any current application. This solution may be useful only for long-term secrets or if a technological jump allows for a great improvement in computation capabilities. 

AEGIS~\cite{wu_aegis_2014} is a lightweight symmetrical algorithm based on the \gls{AES} encryption round function. It can therefore use the \gls{AES} instructions implemented in processors while being twice as fast.

\subsubsection{\gls{UAS} specific research}

In~\cite{avdonin_method_2017}, Avdonin \emph{et al.} propose using \gls{OTP} to secure communication. Beside being the only perfectly secure encryption scheme (in an information theory way of speaking), encryption is performed using only elementary bit-operation \gls{XOR}. This means it is by far the lightest possible solution in terms of computations required. There is a big drawback however: it requires a shared random key, as long as the data transmitted, and usable only once. The memory storage on the drone side would be enormous, but on the server side, with one key per drone it will be multiplied. The solution may still be viable for communication channels which require very little data flow, such as emergency channels. Theoretically, Quantum Key Distribution could provide a secure way to exchange such keys on the fly without resorting to an encryption layer, assuming it is feasible to maintain a quantum communication channel between the ground station and the flying drone. Such assumptions do not seem realistic to the authors, and even by using a classical communication channel with an encryption layer, \gls{OTP} encryption does not seem to be close to optimal in practice.

\subsubsection{Discussion}

Symmetrical algorithms are the simplest way to maintain authenticated communication. They do not use considerably more bandwidth than the cleartext (\gls{AEAD} algorithms can add a few bytes). Chacha20 is old enough to have been thoroughly investigated for faults, and is the more reliable option. If stronger security is required,~\cite{mahdi_improved_2021} may be a better alternative. For systems with hardware acceleration for \gls{AES}, AEGIS is also a good alternative. The key exchange will have to be adapted to the algorithm, as the level of security given to the session key needs to be as high as the symmetrical algorithm's.

If a secure way of transmitting (or storing) very long keys is found however, the \gls{OTP} solution will be by far the fastest and most secure. It is however currently incompatible with the limits of the system.

\section{Securing \gls{UTM} broadcast communications}
\label{sec:broadcast}

\subsection{Securing \gls{ADS-B}-like communication}

It has been proposed to add integrity, authentication and confidentiality to the already formatted \gls{ADS-B} packets in order to communicate with \gls{UTM} services. Even for the commercial aviation there have been numerous propositions to secure the system, as even if \gls{ADS-B} is not the primary source of information for critical systems, it is used more and more often for secondary applications.

The authentication framework proposed by Baek \emph{et al.}~\cite{baek_authentication_2013} adds an identity-based signature to secure \gls{ADS-B}, using a trusted third party to generate their secret key. It has been adapted into a three-level hierarchical identity-based signature (HIBS) by Yang \emph{et al.}~\cite{yang_new_2017}. This improvement reduces the runtime of the protocol, both in signature generation and verification by adding a batch verification mechanism to the system. This system has some lightweight properties, but requires modifications to the \gls{ADS-B} protocol to be implemented, or needs to use a different channel to send the signatures. It does not provide confidentiality regarding the identity of the aircraft.

Another approach for an \gls{ADS-B} security solution was proposed by Yang \emph{et al.}~\cite{yang_lhcsas_2017}. This scheme uses a trusted third party to generate secrets and transmit them to relevant parties. It ensures privacy by using Format-preserving, Feistel-based encryption (FFX) on the unique identification of an aircraft to anonymize them. Attackers are then unable to correlate valuable information to any aircraft. The format preserving nature of FFX makes it compatible with current \gls{ADS-B} systems. The scheme also provides integrity and authentication with the TESLA protocol, using a special identifier on 4 additional \gls{ADS-B} packets to transmit the \gls{MAC} and keys instead of standard \gls{ADS-B} information. This solution requires a loose time synchronization, and has a polynomial overhead in terms of bandwidth and computation. The authors then extended their work~\cite{yang_practical_2019}, achieving adaptive TESLA and solving other problems. They also implemented their solution in a real airport environment, demonstrating the feasibility of their solution. This proposition was designed for commercial aviation and is not optimized for the scale and constraints of \gls{UAS}.

\subsection{Cellular networks}

\gls{ADS-B} is not the preferred solution for the \gls{UAV}-\gls{UTM} link. Most proposed alternatives use cellular network as their communication link, either with the \gls{UAV} directly, or through the \gls{GCS}. 4G and 5G networks include security features, including an authenticated key agreement for a symmetric encryption of the traffic. Therefore the vulnerabilities of the \gls{UAV}-\gls{UTM} link would be dependant on the vulnerabilities discovered within the cellular networks. Ahmad \emph{et al.} have compiled the security issues in cellular networks in a survey~\cite{ahmad_security_2019}. They outline that some issues remain even with the latest versions of these networks, and summarize the potential solutions to clear those issues. 

These solutions cannot be implemented everywhere however, as the cellular coverage will not be available for certain missions. Protecting those communications can be done on the transport layer, with TLS for example. This is far from being the best solution as TLS is very demanding in terms of performance, and is more optimized for larger data flows.

\subsection{RemoteID}

RemoteID is a specification defined by the \gls{FAA}~\cite{noauthor_remoteid_nodate} providing the means to integrate \gls{UAV}s in the American airspace. It includes regulations for communication with external entities such as \gls{UTM} services. \gls{UAV}s need to regularly broadcast timestamped messages containing an identifier, position, heading, speed and emergency status. It does not provide any cryptographic means to protect the information, and is therefore vulnerable to the same attacks as \gls{ADS-B}.

In~\cite{brighente_hide_2022}, Brighente, Conti and Sciancalepore propose different solutions to mitigate the privacy problems brought by this system. The authors present a system able to anonymize and mask the geographical data while detecting 94\% of invasions.

\subsection{Discussion}

Current \gls{UTM} protocols like ADS-B and RemoteID are inherently flawed. The need of transmitting cleartext messages limits considerably the possibility for cryptographic solutions. It is possible to anonymize, but there is a need for authentication with the threat of accepting a lot of false information.

Developing new standards for broadcasting and direct communication with native cryptographic solutions added may be the best way to address the security issues. A signature scheme is sufficient for the broadcast messages (from UAS or UTM systems alike), and a hybrid encryption scheme is preferable for direct communication between a specific \gls{UAV} or pilot and the \gls{UTM} systems.

\section{Challenges, open issues and future directions}
\label{sec:challenges}

The safety of air traffic has been a point of focus since the inception of commercial aviation. Unmanned air traffic is inserted in this environment, and therefore seeks to maintain the level of safety existing in the airspace. Communication is a pillar of traffic management, therefore communication security is an important actor of air traffic safety. For unmanned vehicles, communication is even more important than for manned aircraft, as the control is performed remotely through communication means. In order to develop safe traffic management systems, they must be designed with mandatory security requirements.  The \gls{UAS} field presents strict limitations in terms of performance, limiting the use of classic cryptographic solutions found over the internet. In response to this problem, research has developed lightweight solutions, suited to lower performance systems. This survey presented a non exhaustive overview of the research works proposing cryptographic security solutions for \gls{UAS}, with the intention of ensuring air traffic safety. Privacy and traceability issues are not directly related to the safety of air traffic. These issues were presented, and were noted when covered by the research works. However, they were considered secondary to safety in the recommendations for \gls{UTM} security solutions. By studying these works of research, a few findings emerge which indicates several paths for future research into \gls{UTM} security.

The main challenge for the majority of the authentication protocols presented in this study is the lack of upscaling, and real implementation. Choosing these solutions is not currently recommended due to the lack of experience and testing.

The regulatory framework to authorize \gls{BVLOS} flights will at the minimum impose the levels of security detailed in Section~\ref{subsec:security_goals} as they are mandatory for the security of the \gls{UTM}. The privacy issues can also be addressed by the regulation, but are not directly linked to the safety of \gls{UAS} operations. Those issues are however often resolved by providing the solutions for the security requirements of \gls{UTM}. If some anonymity is required for privacy, maintaining the traceability and non-repudiation will be a challenge and additional work will be needed to provide the property.

Though a lot of research has focused on developing security protocols for \gls{UAS}, there has not been a proposition for a complete security architecture. Integrating the relevant cryptographic solutions into a global scheme will be necessary to meet the security goals required for a secure traffic management. An analytic study is necessary to produce a model that will maintain the security requirements within the whole system. It will then be possible to optimize the model in accordance with the constrains of the system and the security levels deemed necessary for each communication. This architecture can be designed with the network structure of \gls{UAS} and \gls{UTM}. By choosing certain algorithms and implementations, it is possible to optimize the system for improved routing in a \gls{FANET}.

Many authentication protocols rely on passwords, but this is not optimized for large scale infrastructures. Public-key architectures will be necessary for dense traffic areas. Their cost in terms of computation is a lot higher, but the benefits in terms of flexibility, reliability, and bandwidth occupation outweigh those costs for large-scale systems. Public-key architectures can be based on signature or asymmetrical encryption. Developing secure authentication frameworks using them is necessary to reach acceptable levels of security for large scale \gls{UTM}. 

Developing post-quantum solutions will be needed against the approaching threat of quantum computing. The \gls{PQC} standards have been chosen by the \gls{NIST} in July 2022, but they are not lightweight. Quantum resistant lightweight algorithms exist, but they need more testing to be approved as secure against modern cryptanalysis.

\gls{UTM} will also need a signature system, as broadcast messages need to be authenticated. This concerns both the emergency broadcast from the \gls{UTM} system as well as any broadcast message by \gls{UAS} during operations (detect and avoid systems, environmental information...).

\gls{UAS} with larger \gls{UAV}s (over 2 kg) are less restricted in terms of performance. They can support more complex software. It can be recommended to embark non-lightweight standards, as they have withstood the test of time unlike their lightweight counterparts. Due to the increased risk for flying over populated areas, stronger, more reliable security solutions are a relevant alternative. When lightweight standards are considered reliably secure however, these systems should also switch to those lighter solutions.

Research into \gls{UAS} communication security is relatively recent due to the rapidly developing field of \gls{UAV}s. This causes a lack of hindsight for developing safe \gls{UTM} systems. It is therefore important to challenge works of research that could quickly be used into real-world implementations. As the field has no set standards, and is developed privately, many works of research will focus on the field in the coming years. The goal of this survey is also to help criticize and evaluate these new propositions, and to help in maintaining the airspace as safe as it currently is despite the increase in unmanned traffic.

\renewcommand{\glsgroupskip}{}
{\small
\printglossary[type=\acronymtype]}
\renewcommand*{\bibfont}{\small}
\printbibliography

@article{rsa,
  title={A method for obtaining digital signatures and public-key cryptosystems},
  author={Rivest, Ronald L and Shamir, Adi and Adleman, Leonard},
  journal={Communications of the ACM},
  volume={21},
  number={2},
  pages={120--126},
  year={1978},
  publisher={ACM New York, NY, USA}
}

@article{elgamal,
  title={A public key cryptosystem and a signature scheme based on discrete logarithms},
  author={ElGamal, Taher},
  journal={IEEE transactions on information theory},
  volume={31},
  number={4},
  pages={469--472},
  year={1985},
  publisher={IEEE}
}

@article{kwon_empirical_2018,
	title = {Empirical Analysis of {MAVLink} Protocol Vulnerability for Attacking Unmanned Aerial Vehicles},
	volume = {6},
	issn = {2169-3536},
	doi = {10.1109/ACCESS.2018.2863237},
	pages = {43203--43212},
	journaltitle = {{IEEE} Access},
	author = {Kwon, Young-Min and Yu, Jaemin and Cho, Byeong-Moon and Eun, Yongsoon and Park, Kyung-Joon},
	date = {2018},
	note = {Conference Name: {IEEE} Access},
}

@article{shakhatreh_unmanned_2019,
	title = {Unmanned Aerial Vehicles ({UAVs}): A Survey on Civil Applications and Key Research Challenges},
	volume = {7},
	issn = {2169-3536},
	doi = {10.1109/ACCESS.2019.2909530},
	shorttitle = {Unmanned Aerial Vehicles ({UAVs})},
	pages = {48572--48634},
	journaltitle = {{IEEE} Access},
	author = {Shakhatreh, Hazim and Sawalmeh, Ahmad H. and Al-Fuqaha, Ala and Dou, Zuochao and Almaita, Eyad and Khalil, Issa and Othman, Noor Shamsiah and Khreishah, Abdallah and Guizani, Mohsen},
	date = {2019},
	note = {Conference Name: {IEEE} Access},
}

@article{roy_chaotic_2018,
	title = {Chaotic Map-Based Anonymous User Authentication Scheme With User Biometrics and Fuzzy Extractor for Crowdsourcing Internet of Things},
	volume = {5},
	issn = {2327-4662},
	url = {https://ieeexplore.ieee.org/document/7945557},
	doi = {10.1109/JIOT.2017.2714179},
	pages = {2884--2895},
	number = {4},
	journaltitle = {{IEEE} Internet of Things Journal},
	author = {Roy, Sandip and Chatterjee, Santanu and Das, Ashok Kumar and Chattopadhyay, Samiran and Kumari, Saru and Jo, Minho},
	date = {2018-08},
	note = {Conference Name: {IEEE} Internet of Things Journal},
}

@article{srinivas_tcalas_2019,
	title = {{TCALAS}: Temporal Credential-Based Anonymous Lightweight Authentication Scheme for Internet of Drones Environment},
	volume = {68},
	issn = {1939-9359},
	url = {https://ieeexplore.ieee.org/document/8693567},
	doi = {10.1109/TVT.2019.2911672},
	shorttitle = {{TCALAS}},
	pages = {6903--6916},
	number = {7},
	journaltitle = {{IEEE} Transactions on Vehicular Technology},
	author = {Srinivas, Jangirala and Das, Ashok Kumar and Kumar, Neeraj and Rodrigues, Joel J. P. C.},
	date = {2019-07},
	note = {Conference Name: {IEEE} Transactions on Vehicular Technology},
}

@article{ali_securing_2020,
	title = {Securing Smart City Surveillance: A Lightweight Authentication Mechanism for Unmanned Vehicles},
	volume = {8},
	issn = {2169-3536},
	doi = {10.1109/ACCESS.2020.2977817},
	shorttitle = {Securing Smart City Surveillance},
	pages = {43711--43724},
	journaltitle = {{IEEE} Access},
	author = {Ali, Zeeshan and Chaudhry, Shehzad Ashraf and Ramzan, Muhammad Sher and Al-Turjman, Fadi},
	date = {2020},
	note = {Conference Name: {IEEE} Access},
}

@article{wazid_design_2019,
	title = {Design and Analysis of Secure Lightweight Remote User Authentication and Key Agreement Scheme in Internet of Drones Deployment},
	volume = {6},
	issn = {2327-4662},
	doi = {10.1109/JIOT.2018.2888821},
	pages = {3572--3584},
	number = {2},
	journaltitle = {{IEEE} Internet of Things Journal},
	author = {Wazid, Mohammad and Das, Ashok Kumar and Kumar, Neeraj and Vasilakos, Athanasios V. and Rodrigues, Joel J. P. C.},
	date = {2019-04},
	note = {Conference Name: {IEEE} Internet of Things Journal},
}

@article{challa_efficient_2018,
	title = {An efficient {ECC}-based provably secure three-factor user authentication and key agreement protocol for wireless healthcare sensor networks},
	volume = {69},
	issn = {0045-7906},
	url = {https://www.sciencedirect.com/science/article/pii/S0045790616302622},
	doi = {10.1016/j.compeleceng.2017.08.003},
	pages = {534--554},
	journaltitle = {Computers \& Electrical Engineering},
	shortjournal = {Computers \& Electrical Engineering},
	author = {Challa, Sravani and Das, Ashok Kumar and Odelu, Vanga and Kumar, Neeraj and Kumari, Saru and Khan, Muhammad Khurram and Vasilakos, Athanasios V.},
	urldate = {2022-02-23},
	date = {2018-07-01},
	langid = {english},
}

@article{tanveer_lake-iod_2020,
	title = {{LAKE}-{IoD}: Lightweight Authenticated Key Exchange Protocol for the Internet of Drone Environment},
	volume = {8},
	issn = {2169-3536},
	url = {https://ieeexplore.ieee.org/document/9176990},
	doi = {10.1109/ACCESS.2020.3019367},
	shorttitle = {{LAKE}-{IoD}},
	pages = {155645--155659},
	journaltitle = {{IEEE} Access},
	author = {Tanveer, Muhammad and Zahid, Amjad Hussain and Ahmad, Musheer and Baz, Abdullah and Alhakami, Hosam},
	date = {2020},
	note = {Conference Name: {IEEE} Access},
}

@inproceedings{pu_lightweight_2020,
	title = {Lightweight Authentication Protocol for Unmanned Aerial Vehicles Using Physical Unclonable Function and Chaotic System},
	doi = {10.1109/LANMAN49260.2020.9153239},
	eventtitle = {2020 {IEEE} International Symposium on Local and Metropolitan Area Networks ({LANMAN}},
	pages = {1--6},
	booktitle = {2020 {IEEE} International Symposium on Local and Metropolitan Area Networks ({LANMAN}},
	author = {Pu, Cong and Li, Yucheng},
	date = {2020-07},
	note = {{ISSN}: 1944-0375},
}

@article{sharma_communication_2020,
	title = {Communication and networking technologies for {UAVs}: A survey},
	volume = {168},
	issn = {1084-8045},
	url = {https://www.sciencedirect.com/science/article/pii/S1084804520302137},
	doi = {10.1016/j.jnca.2020.102739},
	shorttitle = {Communication and networking technologies for {UAVs}},
	pages = {102739},
	journaltitle = {Journal of Network and Computer Applications},
	shortjournal = {Journal of Network and Computer Applications},
	author = {Sharma, Abhishek and Vanjani, Pankhuri and Paliwal, Nikhil and Basnayaka, Chathuranga M. Wijerathna and Jayakody, Dushantha Nalin K. and Wang, Hwang-Cheng and Muthuchidambaranathan, P.},
	urldate = {2022-02-23},
	date = {2020-10-15},
	langid = {english},
}

@inproceedings{suarez-albela_practical_2018,
	title = {A Practical Performance Comparison of {ECC} and {RSA} for Resource-Constrained {IoT} Devices},
	doi = {10.1109/GIOTS.2018.8534575},
	eventtitle = {2018 Global Internet of Things Summit ({GIoTS})},
	pages = {1--6},
	booktitle = {2018 Global Internet of Things Summit ({GIoTS})},
	author = {Suárez-Albela, Manuel and Fernández-Caramés, Tiago M. and Fraga-Lamas, Paula and Castedo, Luis},
	date = {2018-06},
}

@article{cho_sentinel_2020,
	title = {{SENTINEL}: A Secure and Efficient Authentication Framework for Unmanned Aerial Vehicles},
	volume = {10},
	rights = {http://creativecommons.org/licenses/by/3.0/},
	issn = {2076-3417},
	url = {https://www.mdpi.com/2076-3417/10/9/3149},
	doi = {10.3390/app10093149},
	shorttitle = {{SENTINEL}},
	pages = {3149},
	number = {9},
	journaltitle = {Applied Sciences},
	author = {Cho, Geumhwan and Cho, Junsung and Hyun, Sangwon and Kim, Hyoungshick},
	urldate = {2022-02-22},
	date = {2020-04-30},
	langid = {english},
	note = {Number: 9
Publisher: Multidisciplinary Digital Publishing Institute},
}

@article{jan_design_2021,
	title = {Design and Analysis of Lightweight Authentication Protocol for Securing {IoD}},
	volume = {9},
	issn = {2169-3536},
	doi = {10.1109/ACCESS.2021.3076692},
	pages = {69287--69306},
	journaltitle = {{IEEE} Access},
	author = {Jan, Saeed Ullah and Qayum, Fawad and Khan, Habib Ullah},
	date = {2021},
	note = {Conference Name: {IEEE} Access},
}

@inproceedings{avdonin_method_2017,
	location = {Munich, Germany},
	title = {A method of creating perfectly secure data transmission channel between unmanned aerial vehicle and ground control station based on One-Time pads},
	url = {https://ieeexplore.ieee.org/document/8255167},
	doi = {10.1109/ICUMT.2017.8255167},
	eventtitle = {2017 9th International Congress on Ultra Modern Telecommunications and Control Systems and Workshops ({ICUMT})},
	pages = {410--413},
	booktitle = {2017 9th International Congress on Ultra Modern Telecommunications and Control Systems and Workshops ({ICUMT})},
	author = {Avdonin, Ivan and Budko, Marina and Budko, Mikhail and Grozov, Vladimir and Guirik, Alexei},
	date = {2017-11},
	note = {{ISSN}: 2157-023X},
}

@article{sedjelmaci_hierarchical_2018,
	title = {A Hierarchical Detection and Response System to Enhance Security Against Lethal Cyber-Attacks in {UAV} Networks},
	volume = {48},
	issn = {2168-2232},
	url = {https://ieeexplore.ieee.org/document/7890467},
	doi = {10.1109/TSMC.2017.2681698},
	pages = {1594--1606},
	number = {9},
	journaltitle = {{IEEE} Transactions on Systems, Man, and Cybernetics: Systems},
	author = {Sedjelmaci, Hichem and Senouci, Sidi Mohammed and Ansari, Nirwan},
	date = {2018-09},
	note = {Conference Name: {IEEE} Transactions on Systems, Man, and Cybernetics: Systems},
}

@inproceedings{hooper_securing_2016,
	title = {Securing commercial {WiFi}-based {UAVs} from common security attacks},
	url = {https://ieeexplore.ieee.org/document/7795496},
	doi = {10.1109/MILCOM.2016.7795496},
	eventtitle = {{MILCOM} 2016 - 2016 {IEEE} Military Communications Conference},
	pages = {1213--1218},
	booktitle = {{MILCOM} 2016 - 2016 {IEEE} Military Communications Conference},
	author = {Hooper, Michael and Tian, Yifan and Zhou, Runxuan and Cao, Bin and Lauf, Adrian P. and Watkins, Lanier and Robinson, William H. and Alexis, Wlajimir},
	date = {2016-11},
	note = {{ISSN}: 2155-7586},
}

@article{hayat_survey_2016,
	title = {Survey on Unmanned Aerial Vehicle Networks for Civil Applications: A Communications Viewpoint},
	volume = {18},
	issn = {1553-877X},
	url = {https://ieeexplore.ieee.org/document/7463007},
	doi = {10.1109/COMST.2016.2560343},
	shorttitle = {Survey on Unmanned Aerial Vehicle Networks for Civil Applications},
	pages = {2624--2661},
	number = {4},
	journaltitle = {{IEEE} Communications Surveys Tutorials},
	author = {Hayat, Samira and Yanmaz, Evşen and Muzaffar, Raheeb},
	date = {2016},
	note = {Conference Name: {IEEE} Communications Surveys Tutorials},
}

@article{nikooghadam_provably_2021,
	title = {A provably secure and lightweight authentication scheme for Internet of Drones for smart city surveillance},
	volume = {115},
	issn = {1383-7621},
	url = {https://www.sciencedirect.com/science/article/pii/S138376212030206X},
	doi = {10.1016/j.sysarc.2020.101955},
	pages = {101955},
	journaltitle = {Journal of Systems Architecture},
	shortjournal = {Journal of Systems Architecture},
	author = {Nikooghadam, Mahdi and Amintoosi, Haleh and Islam, {SK} Hafizul and Moghadam, Mostafa Farhadi},
	urldate = {2022-09-07},
	date = {2021-05-01},
	langid = {english},
}

@article{amin_design_2016,
	title = {Design of an anonymity-preserving three-factor authenticated key exchange protocol for wireless sensor networks},
	volume = {101},
	issn = {1389-1286},
	url = {https://www.sciencedirect.com/science/article/pii/S1389128616000207},
	doi = {10.1016/j.comnet.2016.01.006},
	series = {Industrial Technologies and Applications for the Internet of Things},
	pages = {42--62},
	journaltitle = {Computer Networks},
	shortjournal = {Computer Networks},
	author = {Amin, Ruhul and Islam, {SK} Hafizul and Biswas, G. P. and Khan, Muhammad Khurram and Leng, Lu and Kumar, Neeraj},
	urldate = {2022-09-07},
	date = {2016-06-04},
	langid = {english},
}

@article{farash_efficient_2016,
	title = {An efficient user authentication and key agreement scheme for heterogeneous wireless sensor network tailored for the Internet of Things environment},
	volume = {36},
	issn = {1570-8705},
	url = {https://www.sciencedirect.com/science/article/pii/S1570870515001195},
	doi = {10.1016/j.adhoc.2015.05.014},
	pages = {152--176},
	journaltitle = {Ad Hoc Networks},
	shortjournal = {Ad Hoc Networks},
	author = {Farash, Mohammad Sabzinejad and Turkanović, Muhamed and Kumari, Saru and Hölbl, Marko},
	urldate = {2022-09-07},
	date = {2016-01-01},
	langid = {english},
}

@article{zhang_lightweight_2020,
	title = {A lightweight authentication and key agreement scheme for Internet of Drones},
	volume = {154},
	issn = {0140-3664},
	url = {https://www.sciencedirect.com/science/article/pii/S0140366419319358},
	doi = {10.1016/j.comcom.2020.02.067},
	pages = {455--464},
	journaltitle = {Computer Communications},
	shortjournal = {Computer Communications},
	author = {Zhang, Yunru and He, Debiao and Li, Li and Chen, Biwen},
	urldate = {2022-09-07},
	date = {2020-03-15},
	langid = {english},
}

@report{aviaion_rulemaking_committee_unmanned_2022,
	title = {Unmanned Aircraft Systems Beyond Visual Line of Sight},
	url = {https://www.faa.gov/regulations_policies/rulemaking/committees/documents/media/UAS_BVLOS_ARC_FINAL_REPORT_03102022.pdf},
	author = {Aviaion Rulemaking Committee},
	date = {2022-03-10},
}

@report{icao_ads-b_2018,
	title = {{ADS}-B Implementation and operations Guidance Document},
	url = {https://www.icao.int/APAC/Documents/edocs/AIGD%20Edition%2011.pdf},
	author = {{ICAO}},
	urldate = {2022-06-24},
	date = {2018-07},
}

@inproceedings{de_freitas_uav_2010,
	title = {{UAV} relay network to support {WSN} connectivity},
	doi = {10.1109/ICUMT.2010.5676621},
	eventtitle = {International Congress on Ultra Modern Telecommunications and Control Systems},
	pages = {309--314},
	booktitle = {International Congress on Ultra Modern Telecommunications and Control Systems},
	author = {de Freitas, Edison Pignaton and Heimfarth, Tales and Netto, Ivayr Farah and Lino, Carlos Eduardo and Pereira, Carlos Eduardo and Ferreira, Armando Morado and Wagner, Flávio Rech and Larsson, Tony},
	date = {2010-10},
	note = {{ISSN}: 2157-023X},
}

@online{computer_security_division_post-quantum_2017,
	title = {Post-Quantum Cryptography {\textbar} {CSRC} {\textbar} {CSRC}},
	url = {https://csrc.nist.gov/Projects/post-quantum-cryptography},
	titleaddon = {{CSRC} {\textbar} {NIST}},
	author = {Computer Security Division, Information Technology Laboratory},
	urldate = {2022-06-23},
	date = {2017-01-03},
}

@inproceedings{shor_algorithms_1994,
	title = {Algorithms for quantum computation: discrete logarithms and factoring},
	doi = {10.1109/SFCS.1994.365700},
	shorttitle = {Algorithms for quantum computation},
	eventtitle = {Proceedings 35th Annual Symposium on Foundations of Computer Science},
	pages = {124--134},
	booktitle = {Proceedings 35th Annual Symposium on Foundations of Computer Science},
	author = {Shor, P.W.},
	date = {1994-11},
}

@online{computer_security_division_lightweight_2017,
	title = {Lightweight Cryptography {\textbar} {CSRC} {\textbar} {CSRC}},
	url = {https://csrc.nist.gov/projects/lightweight-cryptography},
	titleaddon = {{CSRC} {\textbar} {NIST}},
	author = {Computer Security Division, Information Technology Laboratory},
	urldate = {2022-06-22},
	date = {2017-01-03},
}

@online{noauthor_civil_nodate,
	title = {Civil drones (unmanned aircraft)},
	url = {https://www.easa.europa.eu/domains/civil-drones},
	titleaddon = {{EASA}},
	urldate = {2022-06-21},
	langid = {english},
}

@article{abualigah_applications_2021,
	title = {Applications, Deployments, and Integration of Internet of Drones ({IoD}): A Review},
	volume = {21},
	issn = {1558-1748},
	doi = {10.1109/JSEN.2021.3114266},
	shorttitle = {Applications, Deployments, and Integration of Internet of Drones ({IoD})},
	pages = {25532--25546},
	number = {22},
	journaltitle = {{IEEE} Sensors Journal},
	author = {Abualigah, Laith and Diabat, Ali and Sumari, Putra and Gandomi, Amir H.},
	date = {2021-11},
	note = {Conference Name: {IEEE} Sensors Journal},
}

@online{noauthor_manet_nodate,
	title = {{MANET} vs {VANET} vs {FANET}-Difference between {MANET},{VANET},{FANET}},
	url = {https://www.rfwireless-world.com/Terminology/MANET-vs-VANET-vs-FANET.html},
	urldate = {2022-06-21},
}

@article{zhou_securing_1999,
	title = {Securing ad hoc networks},
	volume = {13},
	issn = {1558-156X},
	doi = {10.1109/65.806983},
	pages = {24--30},
	number = {6},
	journaltitle = {{IEEE} Network},
	author = {Zhou, Lidong and Haas, Z.J.},
	date = {1999-11},
	note = {Conference Name: {IEEE} Network},
}

@article{chriki_fanet_2019,
	title = {{FANET}: Communication, mobility models and security issues},
	volume = {163},
	issn = {1389-1286},
	url = {https://www.sciencedirect.com/science/article/pii/S1389128618309034},
	doi = {10.1016/j.comnet.2019.106877},
	shorttitle = {{FANET}},
	pages = {106877},
	journaltitle = {Computer Networks},
	shortjournal = {Computer Networks},
	author = {Chriki, Amira and Touati, Haifa and Snoussi, Hichem and Kamoun, Farouk},
	urldate = {2022-06-21},
	date = {2019-11-09},
	langid = {english},
}

@article{gharibi_internet_2016,
	title = {Internet of Drones},
	volume = {4},
	issn = {2169-3536},
	doi = {10.1109/ACCESS.2016.2537208},
	pages = {1148--1162},
	journaltitle = {{IEEE} Access},
	author = {Gharibi, Mirmojtaba and Boutaba, Raouf and Waslander, Steven L.},
	date = {2016},
	note = {Conference Name: {IEEE} Access},
}

@online{noauthor_open_nodate,
	title = {Open Category - Civil Drones},
	url = {https://www.easa.europa.eu/domains/civil-drones/drones-regulatory-framework-background/open-category-civil-drones},
	titleaddon = {{EASA}},
	urldate = {2022-06-21},
	langid = {english},
}

@misc{nassi_sok_2019,
	title = {{SoK} - Security and Privacy in the Age of Drones: Threats, Challenges, Solution Mechanisms, and Scientific Gaps},
	url = {http://arxiv.org/abs/1903.05155},
	doi = {10.48550/arXiv.1903.05155},
	shorttitle = {{SoK} - Security and Privacy in the Age of Drones},
	number = {{arXiv}:1903.05155},
	publisher = {{arXiv}},
	author = {Nassi, Ben and Shabtai, Asaf and Masuoka, Ryusuke and Elovici, Yuval},
	urldate = {2022-06-21},
	date = {2019-03-12},
	eprinttype = {arxiv},
	eprint = {1903.05155 [cs]},
	note = {Number: {arXiv}:1903.05155},
}

@online{noauthor_easy_nodate,
	title = {Easy Access Rules for Unmanned Aircraft Systems - Revision from September 2021},
	url = {https://www.easa.europa.eu/document-library/easy-access-rules/online-publications/easy-access-rules-unmanned-aircraft-systems},
	titleaddon = {{EASA}},
	urldate = {2022-06-17},
	langid = {english},
}

@report{icao_unmanned_2019,
	title = {Unmanned Aircraft Systems Traffic Management ({UTM}) – A Common Framework with Core Principles for Global Harmonization},
	url = {https://www.icao.int/safety/UA/Documents/UTM%20Framework%20Edition%203.pdf},
	pages = {45},
	author = {{ICAO}},
	date = {2019},
}

@article{turkanovic_novel_2014,
	title = {A novel user authentication and key agreement scheme for heterogeneous ad hoc wireless sensor networks, based on the Internet of Things notion},
	volume = {20},
	issn = {1570-8705},
	url = {https://www.sciencedirect.com/science/article/pii/S157087051400064X},
	doi = {10.1016/j.adhoc.2014.03.009},
	pages = {96--112},
	journaltitle = {Ad Hoc Networks},
	shortjournal = {Ad Hoc Networks},
	author = {Turkanović, Muhamed and Brumen, Boštjan and Hölbl, Marko},
	urldate = {2022-06-15},
	date = {2014-09-01},
	langid = {english},
}

@online{noauthor_ads-b_nodate,
	title = {{ADS}-B Privacy},
	url = {https://www.faa.gov/air_traffic/technology/equipadsb/privacy/},
	type = {template},
	urldate = {2022-06-15},
	langid = {english},
	note = {Last Modified: 2022-05-26T11:56:00-0400},
}

@online{noauthor_nbaa_2021,
	title = {{NBAA}: Privacy, Security Concerns Remain Regarding {UAS} Remote {ID} Final Rule},
	url = {https://nbaa.org/aircraft-operations/emerging-technologies/uas/nbaa-privacy-security-concerns-remain-regarding-uas-remote-id-final-rule/},
	shorttitle = {{NBAA}},
	titleaddon = {{NBAA} - National Business Aviation Association},
	urldate = {2022-06-15},
	date = {2021-01-19},
	langid = {american},
}

@inproceedings{ezuma_micro-uav_2019,
	title = {Micro-{UAV} Detection and Classification from {RF} Fingerprints Using Machine Learning Techniques},
	doi = {10.1109/AERO.2019.8741970},
	eventtitle = {2019 {IEEE} Aerospace Conference},
	pages = {1--13},
	booktitle = {2019 {IEEE} Aerospace Conference},
	author = {Ezuma, Martins and Erden, Fatih and Anjinappa, Chethan Kumar and Ozdemir, Ozgur and Guvenc, Ismail},
	date = {2019-03},
	note = {{ISSN}: 1095-323X},
}

@article{shoufan_drone_2018,
	title = {Drone Pilot Identification by Classifying Radio-Control Signals},
	volume = {13},
	issn = {1556-6021},
	doi = {10.1109/TIFS.2018.2819126},
	pages = {2439--2447},
	number = {10},
	journaltitle = {{IEEE} Transactions on Information Forensics and Security},
	author = {Shoufan, Abdulhadi and Al-Angari, Haitham M. and Sheikh, Muhammad Faraz Afzal and Damiani, Ernesto},
	date = {2018-10},
	note = {Conference Name: {IEEE} Transactions on Information Forensics and Security},
}

@article{mahdi_improved_2021,
	title = {An improved chacha algorithm for securing data on {IoT} devices},
	volume = {3},
	issn = {2523-3971},
	url = {https://doi.org/10.1007/s42452-021-04425-7},
	doi = {10.1007/s42452-021-04425-7},
	pages = {429},
	number = {4},
	journaltitle = {{SN} Applied Sciences},
	shortjournal = {{SN} Appl. Sci.},
	author = {Mahdi, Mohammed Salih and Hassan, Nidaa Falih and Abdul-Majeed, Ghassan H.},
	urldate = {2022-06-14},
	date = {2021-03-06},
	langid = {english},
}

@article{kane_security_2020,
	title = {Security and Performance in {IoT}: A Balancing Act},
	volume = {8},
	issn = {2169-3536},
	doi = {10.1109/ACCESS.2020.3007536},
	shorttitle = {Security and Performance in {IoT}},
	pages = {121969--121986},
	journaltitle = {{IEEE} Access},
	author = {Kane, Luke E. and Chen, Jiaming James and Thomas, Rebecca and Liu, Vicky and Mckague, Matthew},
	date = {2020},
	note = {Conference Name: {IEEE} Access},
}

@article{mccarthy_fundamental_2020,
	title = {Fundamental Elements of an Urban {UTM}},
	volume = {7},
	rights = {http://creativecommons.org/licenses/by/3.0/},
	issn = {2226-4310},
	url = {https://www.mdpi.com/2226-4310/7/7/85},
	doi = {10.3390/aerospace7070085},
	pages = {85},
	number = {7},
	journaltitle = {Aerospace},
	author = {{McCarthy}, Tim and Pforte, Lars and Burke, Rebekah},
	urldate = {2022-06-14},
	date = {2020-07},
	langid = {english},
	note = {Number: 7
Publisher: Multidisciplinary Digital Publishing Institute},
}

@inproceedings{wu_aegis_2014,
	location = {Berlin, Heidelberg},
	title = {{AEGIS}: A Fast Authenticated Encryption Algorithm},
	isbn = {978-3-662-43414-7},
	doi = {10.1007/978-3-662-43414-7_10},
	series = {Lecture Notes in Computer Science},
	shorttitle = {{AEGIS}},
	pages = {185--201},
	booktitle = {Selected Areas in Cryptography -- {SAC} 2013},
	publisher = {Springer},
	author = {Wu, Hongjun and Preneel, Bart},
	editor = {Lange, Tanja and Lauter, Kristin and Lisoněk, Petr},
	date = {2014},
	langid = {english},
}

@article{anthi_supervised_2019,
	title = {A Supervised Intrusion Detection System for Smart Home {IoT} Devices},
	volume = {6},
	issn = {2327-4662},
	doi = {10.1109/JIOT.2019.2926365},
	pages = {9042--9053},
	number = {5},
	journaltitle = {{IEEE} Internet of Things Journal},
	author = {Anthi, Eirini and Williams, Lowri and Słowińska, Małgorzata and Theodorakopoulos, George and Burnap, Pete},
	date = {2019-10},
	note = {Conference Name: {IEEE} Internet of Things Journal},
}

@article{ponsam_survey_2014,
	title = {A Survey on {MANET} Security Challenges, Attacks and its Countermeasures},
	volume = {3},
	pages = {7},
	number = {1},
	author = {Ponsam, J Godwin and Srinivasan, Dr R},
	date = {2014},
	langid = {english},
}

@article{wu_intelligent_2019,
	title = {Intelligent Reflecting Surface Enhanced Wireless Network via Joint Active and Passive Beamforming},
	volume = {18},
	issn = {1558-2248},
	doi = {10.1109/TWC.2019.2936025},
	pages = {5394--5409},
	number = {11},
	journaltitle = {{IEEE} Transactions on Wireless Communications},
	author = {Wu, Qingqing and Zhang, Rui},
	date = {2019-11},
	note = {Conference Name: {IEEE} Transactions on Wireless Communications},
}

@article{zhong_secure_2019,
	title = {Secure {UAV} Communication With Cooperative Jamming and Trajectory Control},
	volume = {23},
	issn = {1558-2558},
	doi = {10.1109/LCOMM.2018.2889062},
	pages = {286--289},
	number = {2},
	journaltitle = {{IEEE} Communications Letters},
	author = {Zhong, Canhui and Yao, Jianping and Xu, Jie},
	date = {2019-02},
	note = {Conference Name: {IEEE} Communications Letters},
}

@article{chen_design_2018,
	title = {Design and {FPGA}-Based Realization of a Chaotic Secure Video Communication System},
	volume = {28},
	issn = {1558-2205},
	doi = {10.1109/TCSVT.2017.2703946},
	pages = {2359--2371},
	number = {9},
	journaltitle = {{IEEE} Transactions on Circuits and Systems for Video Technology},
	author = {Chen, Shikun and Yu, Simin and Lü, Jinhu and Chen, Guanrong and He, Jianbin},
	date = {2018-09},
	note = {Conference Name: {IEEE} Transactions on Circuits and Systems for Video Technology},
}

@inproceedings{baek_authentication_2013,
	title = {An Authentication Framework for Automatic Dependent Surveillance-Broadcast Based on Online/Offline Identity-Based Signature},
	doi = {10.1109/3PGCIC.2013.61},
	eventtitle = {2013 Eighth International Conference on P2P, Parallel, Grid, Cloud and Internet Computing},
	pages = {358--363},
	booktitle = {2013 Eighth International Conference on P2P, Parallel, Grid, Cloud and Internet Computing},
	author = {Baek, Joonsang and Byon, Young-Ji and Hableel, Eman and Al-Qutayri, Mahmoud},
	date = {2013-10},
}

@article{ahmad_security_2019,
	title = {Security for 5G and Beyond},
	volume = {21},
	issn = {1553-877X},
	doi = {10.1109/COMST.2019.2916180},
	pages = {3682--3722},
	number = {4},
	journaltitle = {{IEEE} Communications Surveys \& Tutorials},
	author = {Ahmad, Ijaz and Shahabuddin, Shahriar and Kumar, Tanesh and Okwuibe, Jude and Gurtov, Andrei and Ylianttila, Mika},
	date = {2019},
	note = {Conference Name: {IEEE} Communications Surveys \& Tutorials},
}

@inproceedings{yang_lhcsas_2017,
	title = {{LHCSAS}: A Lightweight and Highly-Compatible Solution for {ADS}-B Security},
	doi = {10.1109/GLOCOM.2017.8254500},
	shorttitle = {{LHCSAS}},
	eventtitle = {{GLOBECOM} 2017 - 2017 {IEEE} Global Communications Conference},
	pages = {1--7},
	booktitle = {{GLOBECOM} 2017 - 2017 {IEEE} Global Communications Conference},
	author = {Yang, Haomiao and Yao, Mingxuan and Xu, Zili and Liu, Baoshu},
	date = {2017-12},
}

@article{yang_practical_2019,
	title = {A Practical and Compatible Cryptographic Solution to {ADS}-B Security},
	volume = {6},
	issn = {2327-4662},
	doi = {10.1109/JIOT.2018.2882633},
	pages = {3322--3334},
	number = {2},
	journaltitle = {{IEEE} Internet of Things Journal},
	author = {Yang, Haomiao and Zhou, Qixian and Yao, Mingxuan and Lu, Rongxing and Li, Hongwei and Zhang, Xiaosong},
	date = {2019-04},
	note = {Conference Name: {IEEE} Internet of Things Journal},
}

@article{yang_new_2017,
	title = {A New {ADS}-B Authentication Framework Based on Efficient Hierarchical Identity-Based Signature with Batch Verification},
	volume = {10},
	issn = {1939-1374},
	doi = {10.1109/TSC.2015.2459709},
	pages = {165--175},
	number = {2},
	journaltitle = {{IEEE} Transactions on Services Computing},
	author = {Yang, Anjia and Tan, Xiao and Baek, Joonsang and Wong, Duncan S.},
	date = {2017-03},
	note = {Conference Name: {IEEE} Transactions on Services Computing},
}

@article{he_efficient_2015,
	title = {An Efficient Identity-Based Conditional Privacy-Preserving Authentication Scheme for Vehicular Ad Hoc Networks},
	volume = {10},
	issn = {1556-6021},
	doi = {10.1109/TIFS.2015.2473820},
	pages = {2681--2691},
	number = {12},
	journaltitle = {{IEEE} Transactions on Information Forensics and Security},
	author = {He, Debiao and Zeadally, Sherali and Xu, Baowen and Huang, Xinyi},
	date = {2015-12},
	note = {Conference Name: {IEEE} Transactions on Information Forensics and Security},
}

@inproceedings{alsoliman_privacy-preserving_2020,
	title = {Privacy-Preserving Authentication Framework for {UAS} Traffic Management Systems},
	doi = {10.1109/CSNet50428.2020.9265534},
	eventtitle = {2020 4th Cyber Security in Networking Conference ({CSNet})},
	pages = {1--8},
	booktitle = {2020 4th Cyber Security in Networking Conference ({CSNet})},
	author = {Alsoliman, Anas and Rabiah, Abdulrahman Bin and Levorato, Marco},
	date = {2020-10},
}

@article{poor_wireless_2017,
	title = {Wireless physical layer security},
	volume = {114},
	url = {https://www.pnas.org/doi/full/10.1073/pnas.1618130114},
	doi = {10.1073/pnas.1618130114},
	pages = {19--26},
	number = {1},
	journaltitle = {Proceedings of the National Academy of Sciences},
	author = {Poor, H. Vincent and Schaefer, Rafael F.},
	urldate = {2022-05-25},
	date = {2017-01-03},
	note = {Publisher: Proceedings of the National Academy of Sciences},
}

@inproceedings{parlin_jamming_2018,
	title = {Jamming of {UAV} remote control systems using software defined radio},
	doi = {10.1109/ICMCIS.2018.8398711},
	eventtitle = {2018 International Conference on Military Communications and Information Systems ({ICMCIS})},
	pages = {1--6},
	booktitle = {2018 International Conference on Military Communications and Information Systems ({ICMCIS})},
	author = {Pärlin, Karel and Alam, Muhammad Mahtab and Le Moullec, Yannick},
	date = {2018-05},
}

@inproceedings{westerlund_drone_2019,
	title = {Drone Hacking with Raspberry-Pi 3 and {WiFi} Pineapple: Security and Privacy Threats for the Internet-of-Things},
	doi = {10.1109/UVS.2019.8658279},
	shorttitle = {Drone Hacking with Raspberry-Pi 3 and {WiFi} Pineapple},
	eventtitle = {2019 1st International Conference on Unmanned Vehicle Systems-Oman ({UVS})},
	pages = {1--10},
	booktitle = {2019 1st International Conference on Unmanned Vehicle Systems-Oman ({UVS})},
	author = {Westerlund, Ottilia and Asif, Rameez},
	date = {2019-02},
}

@article{ezuma_detection_2020,
	title = {Detection and Classification of {UAVs} Using {RF} Fingerprints in the Presence of Wi-Fi and Bluetooth Interference},
	volume = {1},
	issn = {2644-125X},
	doi = {10.1109/OJCOMS.2019.2955889},
	pages = {60--76},
	journaltitle = {{IEEE} Open Journal of the Communications Society},
	author = {Ezuma, Martins and Erden, Fatih and Kumar Anjinappa, Chethan and Ozdemir, Ozgur and Guvenc, Ismail},
	date = {2020},
	note = {Conference Name: {IEEE} Open Journal of the Communications Society},
}

@report{rudo_consumer_2020,
	title = {Consumer {UAV} Cybersecurity Vulnerability Assessment Using Fuzzing Tests},
	url = {http://arxiv.org/abs/2008.03621},
	number = {{arXiv}:2008.03621},
	institution = {{arXiv}},
	author = {Rudo, David and Zeng, Dr Kai},
	urldate = {2022-05-25},
	date = {2020-08-08},
	doi = {10.48550/arXiv.2008.03621},
	eprinttype = {arxiv},
	eprint = {2008.03621 [cs]},
	note = {type: article},
}

@article{he_drone-assisted_2017,
	title = {Drone-Assisted Public Safety Networks: The Security Aspect},
	volume = {55},
	issn = {1558-1896},
	doi = {10.1109/MCOM.2017.1600799CM},
	shorttitle = {Drone-Assisted Public Safety Networks},
	pages = {218--223},
	number = {8},
	journaltitle = {{IEEE} Communications Magazine},
	author = {He, Daojing and Chan, Sammy and Guizani, Mohsen},
	date = {2017-08},
	note = {Conference Name: {IEEE} Communications Magazine},
}

@article{pirayesh_jamming_2022,
	title = {Jamming Attacks and Anti-Jamming Strategies in Wireless Networks: A Comprehensive Survey},
	volume = {24},
	issn = {1553-877X},
	doi = {10.1109/COMST.2022.3159185},
	shorttitle = {Jamming Attacks and Anti-Jamming Strategies in Wireless Networks},
	pages = {767--809},
	number = {2},
	journaltitle = {{IEEE} Communications Surveys Tutorials},
	author = {Pirayesh, Hossein and Zeng, Huacheng},
	date = {2022},
	note = {Conference Name: {IEEE} Communications Surveys Tutorials},
}

@article{shafique_survey_2021,
	title = {Survey of Security Protocols and Vulnerabilities in Unmanned Aerial Vehicles},
	volume = {9},
	issn = {2169-3536},
	doi = {10.1109/ACCESS.2021.3066778},
	pages = {46927--46948},
	journaltitle = {{IEEE} Access},
	author = {Shafique, Arslan and Mehmood, Abid and Elhadef, Mourad},
	date = {2021},
	note = {Conference Name: {IEEE} Access},
}

@online{noauthor_remoteid_nodate,
	title = {{RemoteID} Final Rule {\textbar} Federal Aviation Administration},
	url = {https://www.faa.gov/newsroom/remoteid-final-rule},
	urldate = {2022-09-13},
}

@inproceedings{brighente_hide_2022,
	location = {New York, {NY}, {USA}},
	title = {Hide and Seek: Privacy-Preserving and {FAA}-compliant Drones Location Tracing},
	isbn = {978-1-4503-9670-7},
	url = {https://doi.org/10.1145/3538969.3543784},
	doi = {10.1145/3538969.3543784},
	series = {{ARES} '22},
	shorttitle = {Hide and Seek},
	pages = {1--11},
	booktitle = {Proceedings of the 17th International Conference on Availability, Reliability and Security},
	publisher = {Association for Computing Machinery},
	author = {Brighente, Alessandro and Conti, Mauro and Sciancalepore, Savio},
	urldate = {2022-09-13},
	date = {2022-08-23},
}

@misc{grover_fast_1996,
	title = {A fast quantum mechanical algorithm for database search},
	url = {http://arxiv.org/abs/quant-ph/9605043},
	doi = {10.48550/arXiv.quant-ph/9605043},
	number = {{arXiv}:quant-ph/9605043},
	publisher = {{arXiv}},
	author = {Grover, Lov K.},
	urldate = {2022-10-19},
	date = {1996-11-19},
	eprinttype = {arxiv},
	eprint = {quant-ph/9605043},
}

@report{chen_documents_2017,
	title = {Documents - Draft Glossary of terms.docx},
	url = {https://www.icao.int/safety/cargosafety/Documents/Forms/AllItems.aspx},
	author = {Chen, Xuefei},
	urldate = {2022-10-19},
	date = {2017-08-11},
}

@incollection{leurent_collisions_2019,
	title = {From Collisions to Chosen-Prefix Collisions Application to Full {SHA}-1},
	isbn = {978-3-030-17652-5},
	pages = {527--555},
	author = {Leurent, Gaëtan and Peyrin, Thomas},
	date = {2019-04-24},
	doi = {10.1007/978-3-030-17659-4_18},
}

\end{document}